\documentclass[aoas,nameyear,seceqn,dvips]{arximspdf}
\usepackage{epic}
\usepackage{graphicx}

\doi{10.1214/10-AOAS329}
\volume{4}
\issue{2}
\pubyear{2010}
\firstpage{645}
\lastpage{662}

\makeatletter
\newcommand{\superscriptchoice}{}
\newcommand{\eqref}[1]{(\ref{#1})}
\makeatother

\begin{document}
\begin{frontmatter}

\title{Bayesian anomaly detection methods for\\ social networks\protect\thanksref{T1}}
\runtitle{Bayesian anomaly detection methods for social networks}
\thankstext[1]{T1}{Supported by a DIF DTC grant.}

\begin{aug}
\author[A]{\fnms{Nicholas A.} \snm{Heard}\ead[label=e1]{n.heard@imperial.ac.uk}\corref{}},
\author[A]{\fnms{David J.} \snm{Weston}\ead[label=e2]{david.weston@imperial.ac.uk}},\break
\author[A]{\fnms{Kiriaki} \snm{Platanioti}}%
\and
\author[A]{\fnms{David J.} \snm{Hand}\thanksref{t2}}%
\thankstext[2]{t2}{Supported by a Royal Society Research Merit Award.}
\runauthor{Heard, Weston, Platanioti and Hand}
\affiliation{Imperial College London}
\address[A]{N. A. Heard\\
D. J. Weston\\
K. Platanioti\\
D. J. Hand\\
Department of Mathematics\\
Imperial College London\\
South Kensington Campus\\
London SW7 2AZ\\
United Kingdom\\
\printead{e1}\\
\phantom{E-mail: }\printead*{e2}}
\end{aug}

\received{\smonth{5} \syear{2009}}
\revised{\smonth{1} \syear{2010}}

%
\begin{abstract}
Learning the network structure of a large graph is computationally
demanding, and dynamically monitoring the network over time for any
changes in structure threatens to be more challenging still.

This paper presents a two-stage method for anomaly detection in
dynamic graphs: the first stage uses simple, conjugate Bayesian models
for discrete time counting processes to track the pairwise links of
all nodes in the graph to assess normality of behavior; the second
stage applies standard network inference tools on a greatly reduced
subset of potentially anomalous nodes. The utility of the method is
demonstrated on simulated and real data sets.
\end{abstract}

%

\end{frontmatter}

\section{Introduction}
Anomaly detection on graphs of social or communication networks has
important security applications. The definition of a graph anomaly
typically depends on the data and application of interest. Typically
anomaly detection focuses on the connections among the graph's
entities and various methods have been developed for their
analysis. Examples include spectral decompositions
[an area excellently summarized in \citet{vonluxburg07}], scan statistics
[\citet{ScanStatsEnron05}] and random walks
[\citet{pan2004}, \citet{FaloutsosRWR06}]. These methods are generally
computationally demanding when applied to very large networks; also,
in deciding upon which one to use, an explicit choice is being made on
the type of anomaly sought. The interest of this paper is anomaly
detection in large dynamic networks, in a context where in principle
any type of anomaly should be detected. We focus on problems relating
to anomalies in social networks, and present analyses of real and
simulated data from this area. In each case, the network is observed
over a sequence of discrete times, where each observation provides
only a partial view of the full connectivity; a complete view of the
network is provided by the time series as a whole.

The real data come from the European Commission Joint Research
Centre's (JRC) European Media Monitor (EMM) (\url{http://emm.jrc.it}).
EMM is a web intelligence service, providing real-time press and media
summaries to Commission cabinets and services, including a breaking
news and alerting service. This service requires JRC to parse each of
the news documents to extract the relevant information and tag the
story as belonging to a particular topic. For our analysis, JRC
provided 131 weeks of Media Monitor data sourced from a collection of
approved websites, starting from 1st January 2005,
although this period includes a known two-week server downtime at the
end of the first month. The data were extracted from news articles
tagged as being related to terrorist attacks, political unrest and
security. The data we receive are undirected and in a simple list
format showing the date of a reported link and the names of the two
individuals involved.

The simulated data come from the VAST Challenge 2008
(\href{http://www.cs.umd.edu/hcil/VASTchallenge08}{http://www.cs.}
\href{http://www.cs.umd.edu/hcil/VASTchallenge08}{umd.edu/hcil/VASTchallenge08}); we consider the
simulated cell phone data from the Mini Challenge focused in the area
of social network analysis. The cell phone call records cover a
fictional ten-day period on an island, narrowed to 400 unique cell
phones during this period. As well as the time of each phone call and
details of who phoned whom, an identifier of the cell tower from which
the call originated is also given. The records should provide critical
information about an important social network structure. From the
results of award winning published work on this challenge
[\citet{yevast2008}], work which used a combination of
\textit{PageRank} [\citet{brinpage1998}] and visual analytic methods,
there is good reason to suspect that the major anomalous activity
occurs on the eighth day and involves a list of at least eleven
individuals.

\section{Two-stage approach}
The idea behind the method presented here is a simple one: If a social
network has fundamentally changed in some important way, then in most
contexts this is likely to suggest that there are some individuals who
are now either communicating more or less \textit{frequently} than
usual, or communicating with \textit{different} individuals than
usual. Beyond this view, there may well be much more subtle network
structure to examine, but initially taking this more simple view
allows good targets to be quickly identified, with the important
possibility to then zoom-in and investigate such local structure.

In this paper we present a two-stage approach to dynamic anomaly
detection. The first stage is a sweep of the database to identify
potentially anomalous nodes in the network; in the second stage, a
subgraph is constructed around this set of nodes, usually extended to
include other nodes which have recently (or, perhaps, ever) communicated
with a node in this set, and then standard network analytic tools are
used to investigate structure in this vastly reduced
subnetwork.\looseness=1

Technically, for each pair of individuals we independently model the
communications between them over time as a counting process, with the
increments of the process following a Bayesian probability model. At
any point in time, we test whether their relationship has changed to a
degree that is statistically significant. If the derived predictive
$p$-value falls below a fixed threshold, this represents a departure
from previously modeled behavior. The node pair are then said to be
anomalous and are added to the set of anomalous nodes for this
period. Such an approach is statistically principled and
computationally very simple. By assuming independence of the
processes, the method is also fully parallelizable, in the sense that
each node pair is examined in isolation. This assumption of
independence will be approximately acceptable only in some
circumstances, and a method which seeks to relax this assumption is
considered in Section \ref{sec:multinomial}.

Once a reduced subset of interesting nodes has been identified,
standard network tools such as spectral clustering can be much more
readily deployed; also, at this stage we are now interested in the
simpler problem of characterizing structure, such as identifying
clusters, rather than looking for changes in this structure, the
latter being a task which requires additional metrics to be specified.

The threshold at which $p$-values are judged to be significant must be
set by the user. In this paper we use a 0.05 threshold, but smaller or
larger critical values would lead to correspondingly smaller or larger
networks of potentially anomalous nodes. In practice, a good threshold
can be chosen to be as large as possible subject to the resulting
anomaly network being of a manageable size such that follow-up
investigation is feasible.

\section{Discrete time counting process models}
The number of communications over time are treated as simple Bayesian
discrete time counting processes with conditionally independent
increments. For each period in time, the number of communications
between individuals will represent the current weight of their
association in the network.

We first consider some different ways of counting up the
communications. Then, simple Bayesian probability models are given for
learning about such counting processes. Full details of these
probability models and the parameterizations used are given in
\ref{suppA} [\citet{Heard10}].

\subsection{Pairwise, individual and total activity analysis}
For each pair of individuals $(i,j)$, starting from time 0 when the
data collection process begins, let $N_{ij}(t)$ be the number of
communications made from $i$ to $j$ up until discrete time $t$;
alternatively, for a simpler binary view of the network, let
$N_{ij}(t)$ be the number of periods in which $i$ has communicated
with $j$ by time $t$. If the graph is undirected, we have the
simplification $N_{ij}(t)\equiv N_{ji}(t)$.

Let $P_{ij}$ be a probability model for the increments $dN_{ij}(t)=
N_{ij}(t)-N_{ij}(t-1)$ under normal circumstances. In the simplest
setting, we can consider $dN_{ij}(1), dN_{ij}(2),\ldots$ as
independent realizations from $P_{ij}$; the distribution $P_{ij}$
corresponds to the normal mode of communication behavior for this
pair of individuals. Anomalous behavior at time $t$, on the other
hand, can be regarded as a value of $dN_{ij}(t)$ drawn from a
distribution other than $P_{ij}$. The aim is then to detect which
values of $dN_{ij}(t)$ are not draws from the unknown $P_{ij}$.

For a realized value of $dN_{ij}(t)=n$, we find a two-sided Bayesian
$p$-value as the posterior probability of observing a count as extreme
as $n$; this posterior distribution is a marginal calculation based on
our revised beliefs about the unknown distribution $P_{ij}$ in light
of all other periods of data we have observed. Carefully chosen
conjugate Bayesian models allow for this inferential process to be
analytically tractable [see \citet{bernardo94} for details]. For
example, a (simplistic) parametric choice for $P_{ij}$ could be
$\operatorname{Poisson}(\lambda_{ij})$ for unknown rate parameter
$\lambda_{ij}>0$. Completing the model specification with a gamma
prior for $\lambda_{ij}$ ensures the posterior predictive distribution
for a future period is calculable as a simple ratio of Poisson--gamma
mass functions. Where no obvious parametric form for~$P_{ij}$ exists,
nonparametric Bayesian inference is available via the Dirichlet
process [\citet{ferg73}].

In the absence of specific prior information about any of the nodes,
identical prior distributions are adopted for each of the node pair
counting processes $N_{ij}(t)$. So in the first observation period,
each node pair has the same probability of being active and, hence, the
implied model on the whole network belongs to the well-known class of
exponential random graph models (ERGMs)
[\citet{wassermanpattison1996}]. From the second time period onward,
however, the posterior predictive distributions will differ between
node pairs according to the activity which has been observed and so
here we see a departure from ERGMs.

The framework above can be regarded as an independent,
\textit{pairwise} analysis of the members of the network. If
$dN_{ij}(t)$ is the \textit{adjacency} of node $i$ to node~$j$ at time~$t$,
then a similar \textit{individual}-based analysis considers the
\textit{outdegree} and \textit{indegree} of node $i$, given by the
respective increments of
%
\begin{eqnarray}
N_{i\cdot}(t)&=&\sum_{j\neq i}N_{ij}(t),\label{eq:Ni.}\\
N_{\cdot i}(t)&=&\sum_{j\neq i}N_{ji}(t)\label{eq:N.i}.
\end{eqnarray}
These two summed processes correspond to the number of outgoing
[Equation~\eqref{eq:Ni.}] and incoming [Equation \eqref{eq:N.i}]
communications over time for individual~$i$. For an undirected graph
the indegree and outdegree are equivalent and Equation \eqref{eq:N.i}
is redundant. Again, we can assume exchangeable increments following
similar probability models for these processes and look for outlying
values in each time period.

Finally, as a highest level summary, we can monitor the \textit{degree
sum} of the network over time, given by the increments of
\[
N_{\cdot\cdot}(t)=\sum_{i} \sum_{j>i}
N_{ij}(t)\quad \mbox{or}\quad N_{\cdot\cdot}(t)=\sum_{i}
\sum_{j\neq i} N_{ij}(t),\label{eq:N}\
\]
where the two definitions correspond to undirected and directed graphs
respectively. Such processes monitor the overall network activity
level. Again, the same conjugate Bayesian probability models can be
applied at this level.

\subsection{Parametric inference and hurdle models}
Social network graphs are typically sparse
[\citet{faloutsos2004}]. Particularly in larger networks, most pairs of
individuals will not communicate with one another, suggesting a
vanishing fraction of node pairs actually have an edge between
them. This sparsity can be seen as providing an analytical advantage
here, as there will be fewer nontrivial node pair relationships in
the graph.

However, when the network is viewed temporally, the sparsity of the
network is further increased. As we will see in the examples later,
even individuals who are related will spend much of the time not
communicating. This type of sparsity is problematic when modeling the
counting processes, as the large number of time periods showing zero
communications mean that standard exponential family distributions are
inappropriate for modeling normal behavior.

We extend the exponential family probability models to their
\textit{hurdle} variants [\citet{mullahy86}], which incorporate
additional probability variables for determining whether or not the
node pair are active in a given period $t$. The modeling of the
process $dN_{ij}(t)$ is split into two parts, first a hurdle process
for determining whether $dN_{ij}(t)=0$ or $dN_{ij}(t)>0$, and then
second another stochastic process governing the value taken by
$dN_{ij}(t)$ at those times when the hurdle process dictates that
$dN_{ij}(t)>0$.

At time $t$ let $A_{ij}(t)$ be the number of time periods $u\leq t$ in
which $dN_{ij}(u)>0$, meaning the node pair $(i,j)$ were active. The
increment for time $t$, $dA_{ij}(t)=A_{ij}(t)-A_{ij}(t-1)$ takes value
0 or 1, with $dA_{ij}(t)=1$ indicating the pair were active in time
period $t$. A counting process model with Bernoulli increments
specifies $A_{ij}(t)$.

For times when the two individuals are active, the hurdle model also
requires a second model for the increments $dN_{ij}(t)\geq1$. We use
the shifted quantities $dN_{ij}(t)-1\geq0$ to define the increments
of a second counting process $dB_{ij}(s)$ by the equations
\begin{eqnarray*}
dB_{ij}(s) &=& dN_{ij}(t_s)-1,\qquad s=1,2,3,\ldots,\\
t_s &=& \min\{t\dvtx A_{ij}(t)=s\},
\end{eqnarray*}
with resulting counting process $B_{ij}(s)=\sum_{u=1}^s dB_{ij}(u)$.

For the hurdle model we therefore need to specify two (typically
independent) models for the counting processes $B_{ij}(\cdot)$ and
$A_{ij}(\cdot)$. Assuming independence of~$A_{ij}(\cdot)$ and
$B_{ij}(\cdot)$, the increments of the compensator
$\Lambda_{ij}(\cdot)$ for the process~$N_{ij}(\cdot)$ can be expressed
as
\[
d\Lambda_{ij}(t)=\mathbb{E}[dN_{ij}(t)|\mathcal{H}_{t-1}]=\mathbb
{E}[dA_{ij}(t)|\mathcal{H}_{t-1}]\bigl(\mathbb{E}[dB_{ij}(t)|\mathcal{H}_{t-1}]+1\bigr),
\]
where for $N(t)=\{N_{ij}(t)\dvtx i\neq j\}$, $\mathcal{H}_{t}=\{N(u)|u=0,1,2,\ldots,t\}$
is the history of the processes up
until time $t$. Then, since
\begin{eqnarray*}
\mathbb{E}[dN_{ij}^2(t)|\mathcal{H}_{t-1}]
&=&\mathbb{E}[dA_{ij}(t)|\mathcal{H}_{t-1}]\bigl(\mathbb
{E}\bigl[\bigl(dB_{ij}(t)+1\bigr)^2|\mathcal{H}_{t-1}\bigr]\bigr)\\
&=&\mathbb{E}[dA_{ij}(t)|\mathcal{H}_{t-1}]\{\operatorname
{var}[dB_{ij}(t)|\mathcal{H}_{t-1}]+1\}\\
&&{}  +d\Lambda_{ij}(t)\mathbb{E}[dB_{ij}(t)|\mathcal{H}_{t-1}],
\end{eqnarray*}
it follows that the increments of the predictable variation of the
counting process martingale $M_{ij}(t)=N_{ij}(t)-\Lambda_{ij}(t)$
satisfy
\begin{eqnarray*}
d\langle M_{ij}(t)\rangle
&=&\mathbb{E}[dA_{ij}(t)|\mathcal{H}_{t-1}]\{\operatorname
{var}[dB_{ij}(t)|\mathcal{H}_{t-1}]+1\}\\
&&{}  +d\Lambda_{ij}(t)\mathbb{E}[dB_{ij}(t)|\mathcal{H}_{t-1}]-d\Lambda_{ij}^2(t).
\end{eqnarray*}
These equations can be used for checking how well the models for
$N_{ij}(t)$ compare in fitting the data.

\subsubsection{Bernoulli process}
The hurdle process increments $\{dA_{ij}(t)\}$ are most simply treated
as a Bernoulli process
\[
dA_{ij}(t)\sim\operatorname{Bernoulli}(\pi_{ij}),\qquad t=1,2,3,\ldots,
\]
where $1-\pi_{ij}$ can now be much greater than the zero count
probability prescribed by standard exponential family models. Note
that this assumes independence of the increments.

\subsubsection{Markov chain}
To enable simple dependence on the activity status in the previous
time period, an alternative Markov model instead considers
%
\begin{eqnarray}
\phi_{ij}&=&\operatorname{Pr}\bigl(dA_{ij}(t)=1|dA_{ij}(t-1)=1\bigr),\label
{eq:markov_p11}\\
\psi_{ij}&=&\operatorname{Pr}\bigl(dA_{ij}(t)=1|dA_{ij}(t-1)=0\bigr).\label
{eq:markov_p01}
\end{eqnarray}

For a comparable marginal probability to $\pi_{ij}$ in the Bernoulli
process model, note that the stationary distribution for this Markov
chain implies an equilibrium probability for the pair $(i,j)$ being
active [$dA_{ij}(t)=1$] at any particular time $t$ given by
\[
\frac{\psi_{ij}}{1+\psi_{ij}-\phi_{ij}}.
\]

Model specification for $B_{ij}$ can use standard exponential family
distributions such as Poisson or geometric; combined with conjugate
beta priors for the hurdle probabilities above, we retain fully
conjugate Bayesian inference for $N_{ij}(t)$.

\subsection{Nonparametric inference}
If, even with the hurdle extensions, it is still unclear what would be
a suitably simple parametric model for the number of communications,
then a useful conjugate, nonparametric Bayesian alternative is the
Dirichlet process (DP) of \citet{ferg73}. Using a base measure which is
a small positive scalar multiple of, say, a hurdle exponential family
distribution, allows fully coherent but data driven inference which is
largely reliant on the tail probabilities of the empirical histogram
of observed counts.

\subsection{Multinomial extensions}\label{sec:multinomial}

For directed graphs, a related approach considers using the ideas
above to first model the overall counting process of activity, say,
for an individual $i$. Then, given a particular number of communications
involving individual $i$, we consider categorical modeling of which
classes of communication they will be. The classes could correspond to
other individuals in the network, or if the links are labeled with
categorical types, these classes could be the link types observed.

Suppose $dN_{i\cdot}(t)=n$, so in the $t$\superscriptchoice{th} time
period individual $i$ makes $n$ communications. Concentrating on whom
the communications were with, let $p_{ij}$ be the probability that any
contact made by individual $i$ will be to individual $j$. Then
assuming independence between subsequent communications,
$dN_{ij}(t)\sim\operatorname{Binomial}(n,p_{ij})$. More generally, using the
vector notation
$dN_{i-}(t)=(dN_{i1}(t),\ldots,dN_{i(i-1)}(t),dN_{i(i+1)}(t),\ldots)$,
%
\[
dN_{i-}(t)\sim\operatorname{Multinomial}
(n,p_{i-}).
\]
Standard conjugate Bayesian inference under the multinomial model uses
a Dirichlet prior for the class probabilities; see \citet{bernardo94}.

For a familiar goodness-of-fit hypothesis test of multinomial data, we
could contrast the observed counts $dN_{ij}(t)$ with the expected $n
p_{ij}$ through the familiar likelihood ratio test statistic
\[
2\sum_{j:dN_{ij}(t)>0} dN_{ij}(t) \log\biggl(\frac{dN_{ij}(t)}{n
\mathbb{E}
(p_{ij})}\biggr),
\]
so performing a $\chi^ 2$ significance test. However, such a test
would not incorporate uncertainty in the overall number of
communications, as it is based conditionally on observing
$dN_{i\cdot}(t)=n$. Hence, we obtain an augmented likelihood ratio test
statistic
\[
2\biggl[\sum_{j:dN_{ij}(t)>0} dN_{ij}(t) \log\biggl(\frac
{dN_{ij}(t)}{n \mathbb{E}
(p_{ij})}\biggr) -
\log\bigl\{P_{i\cdot}\bigl(dN_{i\cdot}(t)=n\bigr)\bigr\}\biggr],
\]
which also takes into account the uncertainty in $dN_{i\cdot}(t)$.

In summary, a probability model for the overall counting process
$N_{i\cdot}(t)$ for individual $i$, along with the multinomial model,
specifies joint a distribution for the pairwise counting processes
$\{N_{ij}(t)\}$. The induced dependence of these split counting
processes on one another will depend on the nature of the probability
model for the total number of observations $N_{i\cdot}(t)$; in the
special case where this model is Poisson, the processes will be
independent of one another.

\section{Sequential and retrospective analyses}
Typically data for a dynamic social network will arrive as an online
stream. At each discrete time $t$ we will have two inferential
possibilities. The first is to decide whether the new data at $t$ is
anomalous compared to the previous data gathered, to which we give the
term \textit{sequential} analysis. For sequential analysis at time
$t$, we are concerned with the distribution $\operatorname
{Pr}(dN_{ij}(t)|\mathcal{H}_{t-1})$. The second possibility is to revise our decisions about
all previous periods in light of the new data, to which we give the
term \textit{retrospective} analysis. For retrospective analysis at
time $t$, we are concerned with the distributions
$\{\operatorname{Pr}(dN_{ij}(u)|\mathcal{H}_{t}/N(u))\dvtx 1\leq u \leq t\}$,
where $\mathcal{H}_{t}/N(u)$ represents the history of the processes if their values
at time $u$ were not observed.

The difference between sequential and retrospective analyses is most
pronounced for times near the start of the process. In sequential
analysis, it is unlikely that the earliest time points will be flagged
as being anomalous, since early on there are very few data points with
which to compare the current observation. However, in retrospective
analysis, we can look back to these early time points and now revise
our opinion, in light of all that has been seen since, as to whether
those periods were in fact anomalous.

Retrospective analysis can be seen as the more thorough inferential
tool, as it contains sequential analysis as a special case. Sequential
analysis alone is faster and more immediately relevant. Once the
process has been running for sufficiently long, subsequent
retrospective analyses of a much earlier time point should eventually
converge in opinion, as should the retrospective and sequential
analyses for more recent time points.

\eject
\section{Results}

\subsection{EMM}
\begin{figure}

\includegraphics{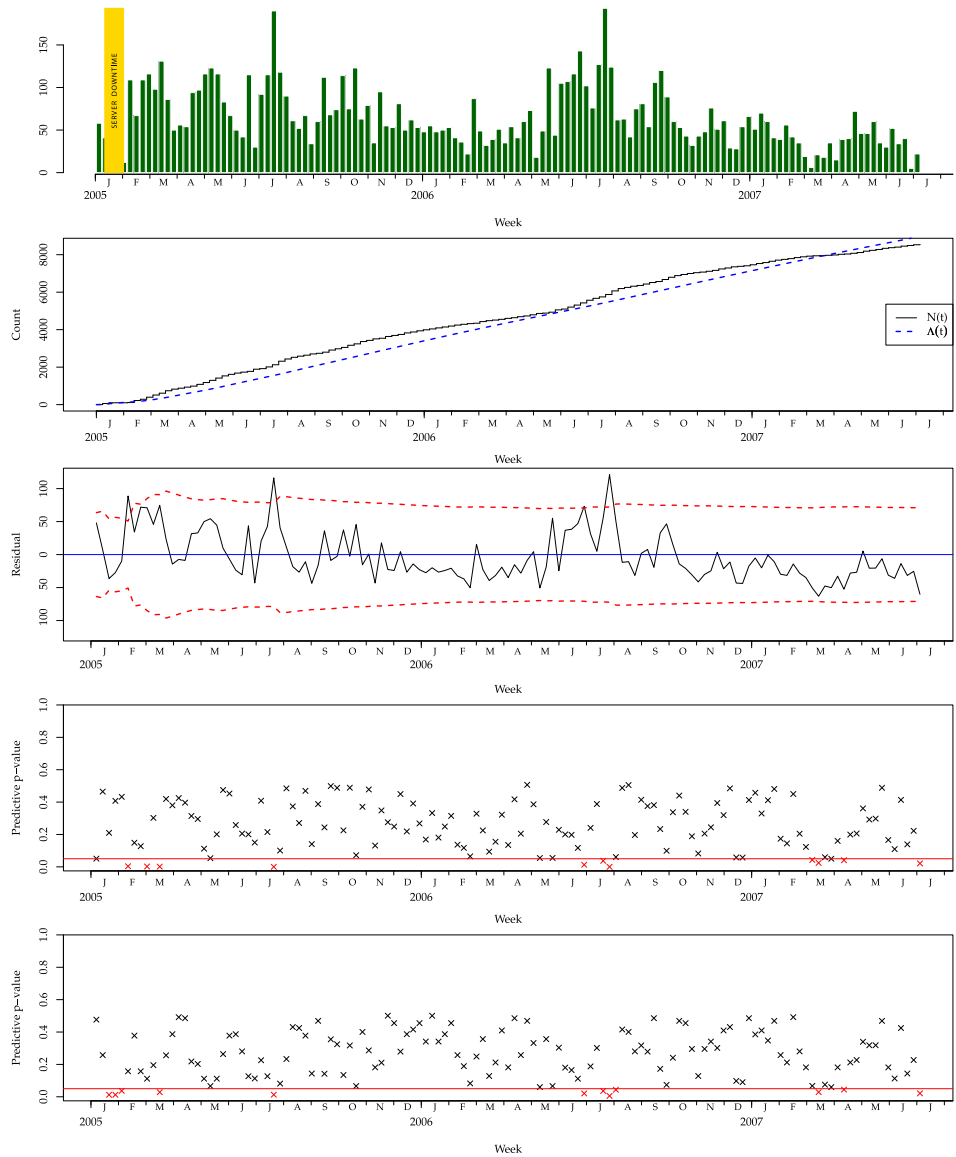}

\caption{Top: The number of contacts made each week by all individuals
in the EMM data set. 2{nd} row: The counting process
and compensator for the whole network activity under the sequential
Dirichlet process model. 3{rd} row: The martingale
residuals; the dashed lines represent 2 times the square root of the
predictable variation of the process. 4{th} row: The
sequential analysis predictive
$p$-values for the observed counts; crosses indicate values falling
below a 0.05 threshold. Bottom: $p$-values from retrospective analysis.}
\label{fig:total_activity_counting_process}
\end{figure}

Here we apply our anomaly detection methods to the real EMM network
data provided by JRC. The weekly counts of the contacts made by all
individuals found in news website stories relating to terrorist
attacks, political unrest or security between 1 January 2005 and 11
July 2007 are shown in the top panel of Figure \ref
{fig:total_activity_counting_process}.

The counting process and compensator for the activity of the whole
network are shown in the second row of Figure
\ref{fig:total_activity_counting_process}. These results have been
obtained from the sequential Dirichlet process model with an
uninformative negative binomial base measure [using parameter pairs
$(0.1,0.01)$ for whole network analysis, and later $(0.1,0.1)$ for
individual and pairwise analyses; see Appendix D of \ref{suppA}, \citet{Heard10}];
parametric analysis with a hurdle Poisson--gamma mixture with the same
parameters gives very similar results. Unsurprisingly, the compensator
is over-predicting activity during the known server downtime occurring
in the first month, and has subsequently under-predicted the total
cumulative activity for a long while after this experience. Note that
the counting process martingale increments and their predictable
variation (third row of Figure~\ref{fig:total_activity_counting_process}) stabilize much earlier than
this, with the only major departures of the residuals from $\pm2$
standard deviations occurring at the corresponding spikes in the count
data. These points also coincide with the lowest predictive $p$-values
in the fourth row of Figure~\ref{fig:total_activity_counting_process}. Note that most of the
remaining significant $p$-values (using a 0.05 threshold) in this graph
correspond to highly negative martingale residuals, suggestive of
further possible server downtimes.

It is instructive to note that the sequential analysis $p$-values do not
show the known server downtime to be significant. This is because we are
still very much in the learning phase when the server failure occurs,
and with uninformative prior beliefs the downtime is quite
acceptable; rather, it is the period immediately after the downtime
that is deemed anomalous. In contrast, a retrospective analysis
(bottom panel of Figure \ref{fig:total_activity_counting_process})
conducted at the end of the study correctly shows the downtime to be
the anomalous period, with very small $p$-values.

For the pairwise and individual analyses we simply discard all data
before the known server downtime. Overall, there are 1814 
individuals involved in the network through the course of the
observation period. The most directly connected individual is the
president of the United States of America during the data collection
period, George W. Bush, eventually making connections with 179 
other nodes. But as mentioned earlier, social network graphs are
typically
sparse and here only 2817 of the possible 1{,}644{,}391 node
connections are ever made.

\begin{figure}

\includegraphics{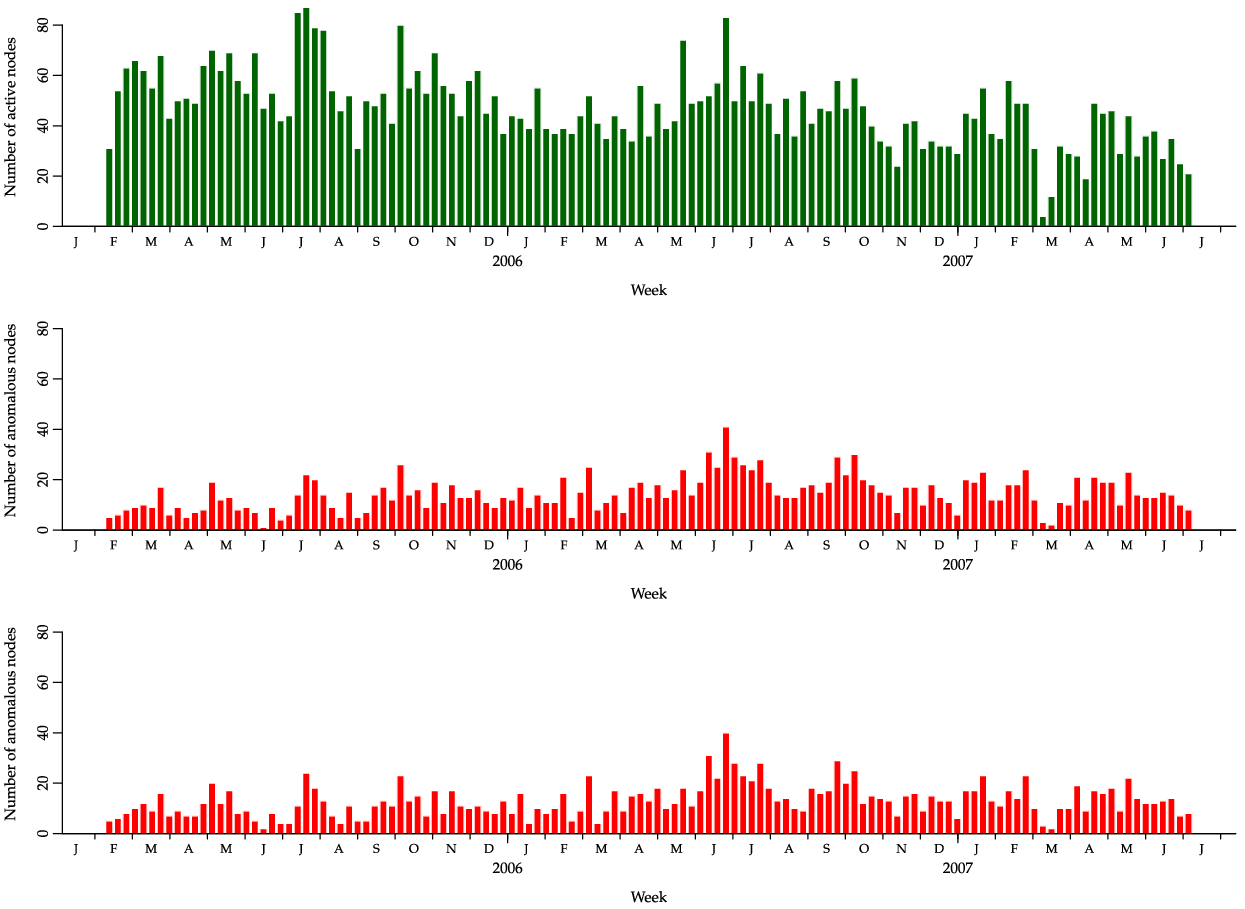}

\caption{The number of active nodes each week (top) and then the
number of anomalous nodes found each week respectively under two
models, the
hurdle Poisson--gamma and the DP with Poisson--gamma base measure.}
\label{fig:anomaly_counts_hpdp}
\end{figure}

\begin{figure}
\fontsize{5}{5.5}\selectfont{\definecolor{lightgrey}{gray}{.85}
\setlength{\unitlength}{1mm}
\vspace*{132pt}\hspace*{-165pt}\begin{picture}(75, 75)
\color{lightgrey}
\dottedline[\href{http://news.google.co.uk/archivesearch?as_q=George+W.+Bush+Junichiro+Koizumi&num=10&hl=en&btnG=Search+Archives&as_epq=&as_oq=&as_eq=&as_user_ldate=2005&as_user_hdate=2007&lr=&as_scoring=t}{$\cdot$}]{3}(91,91)(89,103)
\dottedline[$\bullet$]{0.5}(91,91)(89,103)
\dottedline[\href{http://news.google.co.uk/archivesearch?as_q=George+W.+Bush+Ayman+al-Zawahiri&num=10&hl=en&btnG=Search+Archives&as_epq=&as_oq=&as_eq=&as_user_ldate=2005&as_user_hdate=2007&lr=&as_scoring=t}{$\cdot$}]{3}(91,91)(93,98)
\dottedline[$\bullet$]{0.5}(91,91)(93,98)
\dottedline[\href{http://news.google.co.uk/archivesearch?as_q=George+W.+Bush+Emile+Lahoud&num=10&hl=en&btnG=Search+Archives&as_epq=&as_oq=&as_eq=&as_user_ldate=2005&as_user_hdate=2007&lr=&as_scoring=t}{$\cdot$}]{3}(91,91)(80,89)
\dottedline[$\bullet$]{0.5}(91,91)(80,89)
\dottedline[\href{http://news.google.co.uk/archivesearch?as_q=George+W.+Bush+Norman+Mineta&num=10&hl=en&btnG=Search+Archives&as_epq=&as_oq=&as_eq=&as_user_ldate=2005&as_user_hdate=2007&lr=&as_scoring=t}{$\cdot$}]{3}(91,91)(94,81)
\dottedline[$\bullet$]{0.5}(91,91)(94,81)
\dottedline[\href{http://news.google.co.uk/archivesearch?as_q=Tony+Blair+John+Reid&num=10&hl=en&btnG=Search+Archives&as_epq=&as_oq=&as_eq=&as_user_ldate=2005&as_user_hdate=2007&lr=&as_scoring=t}{$\cdot$}]{3}(13,80)(12,87)
\dottedline[$\bullet$]{0.5}(13,80)(12,87)
\dottedline[\href{http://news.google.co.uk/archivesearch?as_q=Tony+Blair+Vojislav+Koatunica&num=10&hl=en&btnG=Search+Archives&as_epq=&as_oq=&as_eq=&as_user_ldate=2005&as_user_hdate=2007&lr=&as_scoring=t}{$\cdot$}]{3}(13,80)(8,75)
\dottedline[$\bullet$]{0.5}(13,80)(8,75)
\dottedline[\href{http://news.google.co.uk/archivesearch?as_q=Kofi+Annan+Omar+el-Bashir&num=10&hl=en&btnG=Search+Archives&as_epq=&as_oq=&as_eq=&as_user_ldate=2005&as_user_hdate=2007&lr=&as_scoring=t}{$\cdot$}]{3}(65,81)(60,72)
\dottedline[$\bullet$]{0.5}(65,81)(60,72)
\dottedline[\href{http://news.google.co.uk/archivesearch?as_q=Kofi+Annan+Manouchehr+Mottaki&num=10&hl=en&btnG=Search+Archives&as_epq=&as_oq=&as_eq=&as_user_ldate=2005&as_user_hdate=2007&lr=&as_scoring=t}{$\cdot$}]{3}(65,81)(72,70)
\dottedline[$\bullet$]{0.5}(65,81)(72,70)
\dottedline[\href{http://news.google.co.uk/archivesearch?as_q=Dick+Cheney+John+King&num=10&hl=en&btnG=Search+Archives&as_epq=&as_oq=&as_eq=&as_user_ldate=2005&as_user_hdate=2007&lr=&as_scoring=t}{$\cdot$}]{3}(20,65)(24,71)
\dottedline[$\bullet$]{0.5}(20,65)(24,71)
\dottedline[\href{http://news.google.co.uk/archivesearch?as_q=Pervez+Musharraf+Rangin+Dadfar+Spanta&num=10&hl=en&btnG=Search+Archives&as_epq=&as_oq=&as_eq=&as_user_ldate=2005&as_user_hdate=2007&lr=&as_scoring=t}{$\cdot$}]{3}(66,103)(65,111)
\dottedline[$\bullet$]{0.5}(66,103)(65,111)
\dottedline[\href{http://news.google.co.uk/archivesearch?as_q=Junichiro+Koizumi+George+W.+Bush&num=10&hl=en&btnG=Search+Archives&as_epq=&as_oq=&as_eq=&as_user_ldate=2005&as_user_hdate=2007&lr=&as_scoring=t}{$\cdot$}]{3}(89,103)(91,91)
\dottedline[$\bullet$]{0.5}(89,103)(91,91)
\dottedline[\href{http://news.google.co.uk/archivesearch?as_q=Junichiro+Koizumi+Stephen+Harper&num=10&hl=en&btnG=Search+Archives&as_epq=&as_oq=&as_eq=&as_user_ldate=2005&as_user_hdate=2007&lr=&as_scoring=t}{$\cdot$}]{3}(89,103)(88,112)
\dottedline[$\bullet$]{0.5}(89,103)(88,112)
\dottedline[\href{http://news.google.co.uk/archivesearch?as_q=Gordon+Brown+Clare+Short&num=10&hl=en&btnG=Search+Archives&as_epq=&as_oq=&as_eq=&as_user_ldate=2005&as_user_hdate=2007&lr=&as_scoring=t}{$\cdot$}]{3}(115,48)(113,41)
\dottedline[$\bullet$]{0.5}(115,48)(113,41)
\dottedline[\href{http://news.google.co.uk/archivesearch?as_q=Gordon+Brown+Alex+Salmond&num=10&hl=en&btnG=Search+Archives&as_epq=&as_oq=&as_eq=&as_user_ldate=2005&as_user_hdate=2007&lr=&as_scoring=t}{$\cdot$}]{3}(115,48)(119,54)
\dottedline[$\bullet$]{0.5}(115,48)(119,54)
\dottedline[\href{http://news.google.co.uk/archivesearch?as_q=Hamid+Karzai+Ayman+al-Zawahiri&num=10&hl=en&btnG=Search+Archives&as_epq=&as_oq=&as_eq=&as_user_ldate=2005&as_user_hdate=2007&lr=&as_scoring=t}{$\cdot$}]{3}(80,100)(93,98)
\dottedline[$\bullet$]{0.5}(80,100)(93,98)
\dottedline[\href{http://news.google.co.uk/archivesearch?as_q=Hosni+Mubarak+Mahmoud+Abbas&num=10&hl=en&btnG=Search+Archives&as_epq=&as_oq=&as_eq=&as_user_ldate=2005&as_user_hdate=2007&lr=&as_scoring=t}{$\cdot$}]{3}(38,87)(43,96)
\dottedline[$\bullet$]{0.5}(38,87)(43,96)
\dottedline[\href{http://news.google.co.uk/archivesearch?as_q=Hosni+Mubarak+Ali+Larijani&num=10&hl=en&btnG=Search+Archives&as_epq=&as_oq=&as_eq=&as_user_ldate=2005&as_user_hdate=2007&lr=&as_scoring=t}{$\cdot$}]{3}(38,87)(41,80)
\dottedline[$\bullet$]{0.5}(38,87)(41,80)
\dottedline[\href{http://news.google.co.uk/archivesearch?as_q=Javier+Solana+Tzipi+Livni&num=10&hl=en&btnG=Search+Archives&as_epq=&as_oq=&as_eq=&as_user_ldate=2005&as_user_hdate=2007&lr=&as_scoring=t}{$\cdot$}]{3}(49,82)(56,89)
\dottedline[$\bullet$]{0.5}(49,82)(56,89)
\dottedline[\href{http://news.google.co.uk/archivesearch?as_q=Javier+Solana+Ali+Larijani&num=10&hl=en&btnG=Search+Archives&as_epq=&as_oq=&as_eq=&as_user_ldate=2005&as_user_hdate=2007&lr=&as_scoring=t}{$\cdot$}]{3}(49,82)(41,80)
\dottedline[$\bullet$]{0.5}(49,82)(41,80)
\dottedline[\href{http://news.google.co.uk/archivesearch?as_q=John+Reid+Tony+Blair&num=10&hl=en&btnG=Search+Archives&as_epq=&as_oq=&as_eq=&as_user_ldate=2005&as_user_hdate=2007&lr=&as_scoring=t}{$\cdot$}]{3}(12,87)(13,80)
\dottedline[$\bullet$]{0.5}(12,87)(13,80)
\dottedline[\href{http://news.google.co.uk/archivesearch?as_q=Abdullah+G\"{u}l+Manouchehr+Mottaki&num=10&hl=en&btnG=Search+Archives&as_epq=&as_oq=&as_eq=&as_user_ldate=2005&as_user_hdate=2007&lr=&as_scoring=t}{$\cdot$}]{3}(81,64)(72,70)
\dottedline[$\bullet$]{0.5}(81,64)(72,70)
\dottedline[\href{http://news.google.co.uk/archivesearch?as_q=Abdullah+G\"{u}l+Sergey+Kislyak&num=10&hl=en&btnG=Search+Archives&as_epq=&as_oq=&as_eq=&as_user_ldate=2005&as_user_hdate=2007&lr=&as_scoring=t}{$\cdot$}]{3}(81,64)(86,58)
\dottedline[$\bullet$]{0.5}(81,64)(86,58)
\dottedline[\href{http://news.google.co.uk/archivesearch?as_q=Alexander+Downer+Philip+Ruddock&num=10&hl=en&btnG=Search+Archives&as_epq=&as_oq=&as_eq=&as_user_ldate=2005&as_user_hdate=2007&lr=&as_scoring=t}{$\cdot$}]{3}(29,17)(35,20)
\dottedline[$\bullet$]{0.5}(29,17)(35,20)
\dottedline[\href{http://news.google.co.uk/archivesearch?as_q=Vojislav+Koatunica+Tony+Blair&num=10&hl=en&btnG=Search+Archives&as_epq=&as_oq=&as_eq=&as_user_ldate=2005&as_user_hdate=2007&lr=&as_scoring=t}{$\cdot$}]{3}(8,75)(13,80)
\dottedline[$\bullet$]{0.5}(8,75)(13,80)
\dottedline[\href{http://news.google.co.uk/archivesearch?as_q=Ali+Khamenei+Abdoulaye+Wade&num=10&hl=en&btnG=Search+Archives&as_epq=&as_oq=&as_eq=&as_user_ldate=2005&as_user_hdate=2007&lr=&as_scoring=t}{$\cdot$}]{3}(18,96)(20,103)
\dottedline[$\bullet$]{0.5}(18,96)(20,103)
\dottedline[\href{http://news.google.co.uk/archivesearch?as_q=Manmohan+Singh+Mangala+Samaraweera&num=10&hl=en&btnG=Search+Archives&as_epq=&as_oq=&as_eq=&as_user_ldate=2005&as_user_hdate=2007&lr=&as_scoring=t}{$\cdot$}]{3}(86,34)(81,39)
\dottedline[$\bullet$]{0.5}(86,34)(81,39)
\dottedline[\href{http://news.google.co.uk/archivesearch?as_q=Robert+Mueller+Larry+King&num=10&hl=en&btnG=Search+Archives&as_epq=&as_oq=&as_eq=&as_user_ldate=2005&as_user_hdate=2007&lr=&as_scoring=t}{$\cdot$}]{3}(60,24)(66,23)
\dottedline[$\bullet$]{0.5}(60,24)(66,23)
\dottedline[\href{http://news.google.co.uk/archivesearch?as_q=Yoweri+Museveni+Joseph+Kony&num=10&hl=en&btnG=Search+Archives&as_epq=&as_oq=&as_eq=&as_user_ldate=2005&as_user_hdate=2007&lr=&as_scoring=t}{$\cdot$}]{3}(37,47)(37,54)
\dottedline[$\bullet$]{0.5}(37,47)(37,54)
\dottedline[\href{http://news.google.co.uk/archivesearch?as_q=Laura+Bush+L\'{a}szl\'{o}+S\'{o}lyom&num=10&hl=en&btnG=Search+Archives&as_epq=&as_oq=&as_eq=&as_user_ldate=2005&as_user_hdate=2007&lr=&as_scoring=t}{$\cdot$}]{3}(54,119)(47,118)
\dottedline[$\bullet$]{0.5}(54,119)(47,118)
\dottedline[\href{http://news.google.co.uk/archivesearch?as_q=Joseph+Biden+Mark+Warner&num=10&hl=en&btnG=Search+Archives&as_epq=&as_oq=&as_eq=&as_user_ldate=2005&as_user_hdate=2007&lr=&as_scoring=t}{$\cdot$}]{3}(49,34)(50,41)
\dottedline[$\bullet$]{0.5}(49,34)(50,41)
\dottedline[\href{http://news.google.co.uk/archivesearch?as_q=Philip+Ruddock+Alexander+Downer&num=10&hl=en&btnG=Search+Archives&as_epq=&as_oq=&as_eq=&as_user_ldate=2005&as_user_hdate=2007&lr=&as_scoring=t}{$\cdot$}]{3}(35,20)(29,17)
\dottedline[$\bullet$]{0.5}(35,20)(29,17)
\dottedline[\href{http://news.google.co.uk/archivesearch?as_q=Philip+Ruddock+Brendan+Nelson&num=10&hl=en&btnG=Search+Archives&as_epq=&as_oq=&as_eq=&as_user_ldate=2005&as_user_hdate=2007&lr=&as_scoring=t}{$\cdot$}]{3}(35,20)(42,16)
\dottedline[$\bullet$]{0.5}(35,20)(42,16)
\dottedline[\href{http://news.google.co.uk/archivesearch?as_q=Jalal+Talabani+Mahmoud+Othman&num=10&hl=en&btnG=Search+Archives&as_epq=&as_oq=&as_eq=&as_user_ldate=2005&as_user_hdate=2007&lr=&as_scoring=t}{$\cdot$}]{3}(86,17)(91,13)
\dottedline[$\bullet$]{0.5}(86,17)(91,13)
\dottedline[\href{http://news.google.co.uk/archivesearch?as_q=Des+Browne+Clare+Short&num=10&hl=en&btnG=Search+Archives&as_epq=&as_oq=&as_eq=&as_user_ldate=2005&as_user_hdate=2007&lr=&as_scoring=t}{$\cdot$}]{3}(110,35)(113,41)
\dottedline[$\bullet$]{0.5}(110,35)(113,41)
\dottedline[\href{http://news.google.co.uk/archivesearch?as_q=Stephen+Harper+Junichiro+Koizumi&num=10&hl=en&btnG=Search+Archives&as_epq=&as_oq=&as_eq=&as_user_ldate=2005&as_user_hdate=2007&lr=&as_scoring=t}{$\cdot$}]{3}(88,112)(89,103)
\dottedline[$\bullet$]{0.5}(88,112)(89,103)
\dottedline[\href{http://news.google.co.uk/archivesearch?as_q=Omar+el-Bashir+Kofi+Annan&num=10&hl=en&btnG=Search+Archives&as_epq=&as_oq=&as_eq=&as_user_ldate=2005&as_user_hdate=2007&lr=&as_scoring=t}{$\cdot$}]{3}(60,72)(65,81)
\dottedline[$\bullet$]{0.5}(60,72)(65,81)
\dottedline[\href{http://news.google.co.uk/archivesearch?as_q=Omar+el-Bashir+Alpha+Oumar+Konare&num=10&hl=en&btnG=Search+Archives&as_epq=&as_oq=&as_eq=&as_user_ldate=2005&as_user_hdate=2007&lr=&as_scoring=t}{$\cdot$}]{3}(60,72)(55,64)
\dottedline[$\bullet$]{0.5}(60,72)(55,64)
\dottedline[\href{http://news.google.co.uk/archivesearch?as_q=+James+Jones+Brendan+Nelson&num=10&hl=en&btnG=Search+Archives&as_epq=&as_oq=&as_eq=&as_user_ldate=2005&as_user_hdate=2007&lr=&as_scoring=t}{$\cdot$}]{3}(43,9)(42,16)
\dottedline[$\bullet$]{0.5}(43,9)(42,16)
\dottedline[\href{http://news.google.co.uk/archivesearch?as_q=Mahmoud+Abbas+Hosni+Mubarak&num=10&hl=en&btnG=Search+Archives&as_epq=&as_oq=&as_eq=&as_user_ldate=2005&as_user_hdate=2007&lr=&as_scoring=t}{$\cdot$}]{3}(43,96)(38,87)
\dottedline[$\bullet$]{0.5}(43,96)(38,87)
\dottedline[\href{http://news.google.co.uk/archivesearch?as_q=Mahmoud+Abbas+Tzipi+Livni&num=10&hl=en&btnG=Search+Archives&as_epq=&as_oq=&as_eq=&as_user_ldate=2005&as_user_hdate=2007&lr=&as_scoring=t}{$\cdot$}]{3}(43,96)(56,89)
\dottedline[$\bullet$]{0.5}(43,96)(56,89)
\dottedline[\href{http://news.google.co.uk/archivesearch?as_q=Mahmoud+Abbas+Elie+Wiesel&num=10&hl=en&btnG=Search+Archives&as_epq=&as_oq=&as_eq=&as_user_ldate=2005&as_user_hdate=2007&lr=&as_scoring=t}{$\cdot$}]{3}(43,96)(35,102)
\dottedline[$\bullet$]{0.5}(43,96)(35,102)
\dottedline[\href{http://news.google.co.uk/archivesearch?as_q=Mahmoud+Othman+Jalal+Talabani&num=10&hl=en&btnG=Search+Archives&as_epq=&as_oq=&as_eq=&as_user_ldate=2005&as_user_hdate=2007&lr=&as_scoring=t}{$\cdot$}]{3}(91,13)(86,17)
\dottedline[$\bullet$]{0.5}(91,13)(86,17)
\dottedline[\href{http://news.google.co.uk/archivesearch?as_q=Mark+Warner+Joseph+Biden&num=10&hl=en&btnG=Search+Archives&as_epq=&as_oq=&as_eq=&as_user_ldate=2005&as_user_hdate=2007&lr=&as_scoring=t}{$\cdot$}]{3}(50,41)(49,34)
\dottedline[$\bullet$]{0.5}(50,41)(49,34)
\dottedline[\href{http://news.google.co.uk/archivesearch?as_q=Clare+Short+Gordon+Brown&num=10&hl=en&btnG=Search+Archives&as_epq=&as_oq=&as_eq=&as_user_ldate=2005&as_user_hdate=2007&lr=&as_scoring=t}{$\cdot$}]{3}(113,41)(115,48)
\dottedline[$\bullet$]{0.5}(113,41)(115,48)
\dottedline[\href{http://news.google.co.uk/archivesearch?as_q=Clare+Short+Des+Browne&num=10&hl=en&btnG=Search+Archives&as_epq=&as_oq=&as_eq=&as_user_ldate=2005&as_user_hdate=2007&lr=&as_scoring=t}{$\cdot$}]{3}(113,41)(110,35)
\dottedline[$\bullet$]{0.5}(113,41)(110,35)
\dottedline[\href{http://news.google.co.uk/archivesearch?as_q=Tzipi+Livni+Javier+Solana&num=10&hl=en&btnG=Search+Archives&as_epq=&as_oq=&as_eq=&as_user_ldate=2005&as_user_hdate=2007&lr=&as_scoring=t}{$\cdot$}]{3}(56,89)(49,82)
\dottedline[$\bullet$]{0.5}(56,89)(49,82)
\dottedline[\href{http://news.google.co.uk/archivesearch?as_q=Tzipi+Livni+Mahmoud+Abbas&num=10&hl=en&btnG=Search+Archives&as_epq=&as_oq=&as_eq=&as_user_ldate=2005&as_user_hdate=2007&lr=&as_scoring=t}{$\cdot$}]{3}(56,89)(43,96)
\dottedline[$\bullet$]{0.5}(56,89)(43,96)
\dottedline[\href{http://news.google.co.uk/archivesearch?as_q=Abdoulaye+Wade+Ali+Khamenei&num=10&hl=en&btnG=Search+Archives&as_epq=&as_oq=&as_eq=&as_user_ldate=2005&as_user_hdate=2007&lr=&as_scoring=t}{$\cdot$}]{3}(20,103)(18,96)
\dottedline[$\bullet$]{0.5}(20,103)(18,96)
\dottedline[\href{http://news.google.co.uk/archivesearch?as_q=Alpha+Oumar+Konare+Omar+el-Bashir&num=10&hl=en&btnG=Search+Archives&as_epq=&as_oq=&as_eq=&as_user_ldate=2005&as_user_hdate=2007&lr=&as_scoring=t}{$\cdot$}]{3}(55,64)(60,72)
\dottedline[$\bullet$]{0.5}(55,64)(60,72)
\dottedline[\href{http://news.google.co.uk/archivesearch?as_q=Alex+Salmond+Gordon+Brown&num=10&hl=en&btnG=Search+Archives&as_epq=&as_oq=&as_eq=&as_user_ldate=2005&as_user_hdate=2007&lr=&as_scoring=t}{$\cdot$}]{3}(119,54)(115,48)
\dottedline[$\bullet$]{0.5}(119,54)(115,48)
\dottedline[\href{http://news.google.co.uk/archivesearch?as_q=Khaled+Mashal+Haim+Ramon&num=10&hl=en&btnG=Search+Archives&as_epq=&as_oq=&as_eq=&as_user_ldate=2005&as_user_hdate=2007&lr=&as_scoring=t}{$\cdot$}]{3}(20,50)(23,44)
\dottedline[$\bullet$]{0.5}(20,50)(23,44)
\dottedline[\href{http://news.google.co.uk/archivesearch?as_q=Vaughn+Walker+Brad+Berenson&num=10&hl=en&btnG=Search+Archives&as_epq=&as_oq=&as_eq=&as_user_ldate=2005&as_user_hdate=2007&lr=&as_scoring=t}{$\cdot$}]{3}(97,43)(98,49)
\dottedline[$\bullet$]{0.5}(97,43)(98,49)
\dottedline[\href{http://news.google.co.uk/archivesearch?as_q=Frank-Walter+Steinmeier+Manouchehr+Mottaki&num=10&hl=en&btnG=Search+Archives&as_epq=&as_oq=&as_eq=&as_user_ldate=2005&as_user_hdate=2007&lr=&as_scoring=t}{$\cdot$}]{3}(71,61)(72,70)
\dottedline[$\bullet$]{0.5}(71,61)(72,70)
\dottedline[\href{http://news.google.co.uk/archivesearch?as_q=Ayman+al-Zawahiri+George+W.+Bush&num=10&hl=en&btnG=Search+Archives&as_epq=&as_oq=&as_eq=&as_user_ldate=2005&as_user_hdate=2007&lr=&as_scoring=t}{$\cdot$}]{3}(93,98)(91,91)
\dottedline[$\bullet$]{0.5}(93,98)(91,91)
\dottedline[\href{http://news.google.co.uk/archivesearch?as_q=Ayman+al-Zawahiri+Hamid+Karzai&num=10&hl=en&btnG=Search+Archives&as_epq=&as_oq=&as_eq=&as_user_ldate=2005&as_user_hdate=2007&lr=&as_scoring=t}{$\cdot$}]{3}(93,98)(80,100)
\dottedline[$\bullet$]{0.5}(93,98)(80,100)
\dottedline[\href{http://news.google.co.uk/archivesearch?as_q=Ayman+al-Zawahiri+Nouri+al-Maliki&num=10&hl=en&btnG=Search+Archives&as_epq=&as_oq=&as_eq=&as_user_ldate=2005&as_user_hdate=2007&lr=&as_scoring=t}{$\cdot$}]{3}(93,98)(100,105)
\dottedline[$\bullet$]{0.5}(93,98)(100,105)
\dottedline[\href{http://news.google.co.uk/archivesearch?as_q=Emile+Lahoud+George+W.+Bush&num=10&hl=en&btnG=Search+Archives&as_epq=&as_oq=&as_eq=&as_user_ldate=2005&as_user_hdate=2007&lr=&as_scoring=t}{$\cdot$}]{3}(80,89)(91,91)
\dottedline[$\bullet$]{0.5}(80,89)(91,91)
\dottedline[\href{http://news.google.co.uk/archivesearch?as_q=Amir+Peretz+Dan+Halutz&num=10&hl=en&btnG=Search+Archives&as_epq=&as_oq=&as_eq=&as_user_ldate=2005&as_user_hdate=2007&lr=&as_scoring=t}{$\cdot$}]{3}(99,26)(104,23)
\dottedline[$\bullet$]{0.5}(99,26)(104,23)
\dottedline[\href{http://news.google.co.uk/archivesearch?as_q=Haim+Ramon+Khaled+Mashal&num=10&hl=en&btnG=Search+Archives&as_epq=&as_oq=&as_eq=&as_user_ldate=2005&as_user_hdate=2007&lr=&as_scoring=t}{$\cdot$}]{3}(23,44)(20,50)
\dottedline[$\bullet$]{0.5}(23,44)(20,50)
\dottedline[\href{http://news.google.co.uk/archivesearch?as_q=Xanana+Gusm\H{a}o+Mar\'{\i}+Alkatiri&num=10&hl=en&btnG=Search+Archives&as_epq=&as_oq=&as_eq=&as_user_ldate=2005&as_user_hdate=2007&lr=&as_scoring=t}{$\cdot$}]{3}(7,57)(5,64)
\dottedline[$\bullet$]{0.5}(7,57)(5,64)
\dottedline[\href{http://news.google.co.uk/archivesearch?as_q=Mar\'{\i}+Alkatiri+Xanana+Gusm\H{a}o&num=10&hl=en&btnG=Search+Archives&as_epq=&as_oq=&as_eq=&as_user_ldate=2005&as_user_hdate=2007&lr=&as_scoring=t}{$\cdot$}]{3}(5,64)(7,57)
\dottedline[$\bullet$]{0.5}(5,64)(7,57)
\dottedline[\href{http://news.google.co.uk/archivesearch?as_q=Norman+Mineta+George+W.+Bush&num=10&hl=en&btnG=Search+Archives&as_epq=&as_oq=&as_eq=&as_user_ldate=2005&as_user_hdate=2007&lr=&as_scoring=t}{$\cdot$}]{3}(94,81)(91,91)
\dottedline[$\bullet$]{0.5}(94,81)(91,91)
\dottedline[\href{http://news.google.co.uk/archivesearch?as_q=Norman+Mineta+Tony+Snow&num=10&hl=en&btnG=Search+Archives&as_epq=&as_oq=&as_eq=&as_user_ldate=2005&as_user_hdate=2007&lr=&as_scoring=t}{$\cdot$}]{3}(94,81)(98,74)
\dottedline[$\bullet$]{0.5}(94,81)(98,74)
\dottedline[\href{http://news.google.co.uk/archivesearch?as_q=Brendan+Nelson+Philip+Ruddock&num=10&hl=en&btnG=Search+Archives&as_epq=&as_oq=&as_eq=&as_user_ldate=2005&as_user_hdate=2007&lr=&as_scoring=t}{$\cdot$}]{3}(42,16)(35,20)
\dottedline[$\bullet$]{0.5}(42,16)(35,20)
\dottedline[\href{http://news.google.co.uk/archivesearch?as_q=Brendan+Nelson++James+Jones&num=10&hl=en&btnG=Search+Archives&as_epq=&as_oq=&as_eq=&as_user_ldate=2005&as_user_hdate=2007&lr=&as_scoring=t}{$\cdot$}]{3}(42,16)(43,9)
\dottedline[$\bullet$]{0.5}(42,16)(43,9)
\dottedline[\href{http://news.google.co.uk/archivesearch?as_q=David+Cameron+Shami+Chakrabarti&num=10&hl=en&btnG=Search+Archives&as_epq=&as_oq=&as_eq=&as_user_ldate=2005&as_user_hdate=2007&lr=&as_scoring=t}{$\cdot$}]{3}(7,45)(10,38)
\dottedline[$\bullet$]{0.5}(7,45)(10,38)
\dottedline[\href{http://news.google.co.uk/archivesearch?as_q=Joseph+Kony+Yoweri+Museveni&num=10&hl=en&btnG=Search+Archives&as_epq=&as_oq=&as_eq=&as_user_ldate=2005&as_user_hdate=2007&lr=&as_scoring=t}{$\cdot$}]{3}(37,54)(37,47)
\dottedline[$\bullet$]{0.5}(37,54)(37,47)
\dottedline[\href{http://news.google.co.uk/archivesearch?as_q=Joseph+Kony+Riek+Machar&num=10&hl=en&btnG=Search+Archives&as_epq=&as_oq=&as_eq=&as_user_ldate=2005&as_user_hdate=2007&lr=&as_scoring=t}{$\cdot$}]{3}(37,54)(37,61)
\dottedline[$\bullet$]{0.5}(37,54)(37,61)
\dottedline[\href{http://news.google.co.uk/archivesearch?as_q=Shami+Chakrabarti+David+Cameron&num=10&hl=en&btnG=Search+Archives&as_epq=&as_oq=&as_eq=&as_user_ldate=2005&as_user_hdate=2007&lr=&as_scoring=t}{$\cdot$}]{3}(10,38)(7,45)
\dottedline[$\bullet$]{0.5}(10,38)(7,45)
\dottedline[\href{http://news.google.co.uk/archivesearch?as_q=Mangala+Samaraweera+Manmohan+Singh&num=10&hl=en&btnG=Search+Archives&as_epq=&as_oq=&as_eq=&as_user_ldate=2005&as_user_hdate=2007&lr=&as_scoring=t}{$\cdot$}]{3}(81,39)(86,34)
\dottedline[$\bullet$]{0.5}(81,39)(86,34)
\dottedline[\href{http://news.google.co.uk/archivesearch?as_q=Taro+Aso++Thomas+Schieffer&num=10&hl=en&btnG=Search+Archives&as_epq=&as_oq=&as_eq=&as_user_ldate=2005&as_user_hdate=2007&lr=&as_scoring=t}{$\cdot$}]{3}(118,74)(116,80)
\dottedline[$\bullet$]{0.5}(118,74)(116,80)
\dottedline[\href{http://news.google.co.uk/archivesearch?as_q=Larry+King+Robert+Mueller&num=10&hl=en&btnG=Search+Archives&as_epq=&as_oq=&as_eq=&as_user_ldate=2005&as_user_hdate=2007&lr=&as_scoring=t}{$\cdot$}]{3}(66,23)(60,24)
\dottedline[$\bullet$]{0.5}(66,23)(60,24)
\dottedline[\href{http://news.google.co.uk/archivesearch?as_q=Elie+Wiesel+Mahmoud+Abbas&num=10&hl=en&btnG=Search+Archives&as_epq=&as_oq=&as_eq=&as_user_ldate=2005&as_user_hdate=2007&lr=&as_scoring=t}{$\cdot$}]{3}(35,102)(43,96)
\dottedline[$\bullet$]{0.5}(35,102)(43,96)
\dottedline[\href{http://news.google.co.uk/archivesearch?as_q=Dan+Halutz+Amir+Peretz&num=10&hl=en&btnG=Search+Archives&as_epq=&as_oq=&as_eq=&as_user_ldate=2005&as_user_hdate=2007&lr=&as_scoring=t}{$\cdot$}]{3}(104,23)(99,26)
\dottedline[$\bullet$]{0.5}(104,23)(99,26)
\dottedline[\href{http://news.google.co.uk/archivesearch?as_q=John+King+Dick+Cheney&num=10&hl=en&btnG=Search+Archives&as_epq=&as_oq=&as_eq=&as_user_ldate=2005&as_user_hdate=2007&lr=&as_scoring=t}{$\cdot$}]{3}(24,71)(20,65)
\dottedline[$\bullet$]{0.5}(24,71)(20,65)
\dottedline[\href{http://news.google.co.uk/archivesearch?as_q=Omar+Karami+Samir+Geagea&num=10&hl=en&btnG=Search+Archives&as_epq=&as_oq=&as_eq=&as_user_ldate=2005&as_user_hdate=2007&lr=&as_scoring=t}{$\cdot$}]{3}(76,7)(70,7)
\dottedline[$\bullet$]{0.5}(76,7)(70,7)
\dottedline[\href{http://news.google.co.uk/archivesearch?as_q=Manouchehr+Mottaki+Kofi+Annan&num=10&hl=en&btnG=Search+Archives&as_epq=&as_oq=&as_eq=&as_user_ldate=2005&as_user_hdate=2007&lr=&as_scoring=t}{$\cdot$}]{3}(72,70)(65,81)
\dottedline[$\bullet$]{0.5}(72,70)(65,81)
\dottedline[\href{http://news.google.co.uk/archivesearch?as_q=Manouchehr+Mottaki+Abdullah+G\"{u}l&num=10&hl=en&btnG=Search+Archives&as_epq=&as_oq=&as_eq=&as_user_ldate=2005&as_user_hdate=2007&lr=&as_scoring=t}{$\cdot$}]{3}(72,70)(81,64)
\dottedline[$\bullet$]{0.5}(72,70)(81,64)
\dottedline[\href{http://news.google.co.uk/archivesearch?as_q=Manouchehr+Mottaki+Frank-Walter+Steinmeier&num=10&hl=en&btnG=Search+Archives&as_epq=&as_oq=&as_eq=&as_user_ldate=2005&as_user_hdate=2007&lr=&as_scoring=t}{$\cdot$}]{3}(72,70)(71,61)
\dottedline[$\bullet$]{0.5}(72,70)(71,61)
\dottedline[\href{http://news.google.co.uk/archivesearch?as_q=Nouri+al-Maliki+Ayman+al-Zawahiri&num=10&hl=en&btnG=Search+Archives&as_epq=&as_oq=&as_eq=&as_user_ldate=2005&as_user_hdate=2007&lr=&as_scoring=t}{$\cdot$}]{3}(100,105)(93,98)
\dottedline[$\bullet$]{0.5}(100,105)(93,98)
\dottedline[\href{http://news.google.co.uk/archivesearch?as_q=Ali+Larijani+Hosni+Mubarak&num=10&hl=en&btnG=Search+Archives&as_epq=&as_oq=&as_eq=&as_user_ldate=2005&as_user_hdate=2007&lr=&as_scoring=t}{$\cdot$}]{3}(41,80)(38,87)
\dottedline[$\bullet$]{0.5}(41,80)(38,87)
\dottedline[\href{http://news.google.co.uk/archivesearch?as_q=Ali+Larijani+Javier+Solana&num=10&hl=en&btnG=Search+Archives&as_epq=&as_oq=&as_eq=&as_user_ldate=2005&as_user_hdate=2007&lr=&as_scoring=t}{$\cdot$}]{3}(41,80)(49,82)
\dottedline[$\bullet$]{0.5}(41,80)(49,82)
\dottedline[\href{http://news.google.co.uk/archivesearch?as_q=Sergey+Kislyak+Abdullah+G\"{u}l&num=10&hl=en&btnG=Search+Archives&as_epq=&as_oq=&as_eq=&as_user_ldate=2005&as_user_hdate=2007&lr=&as_scoring=t}{$\cdot$}]{3}(86,58)(81,64)
\dottedline[$\bullet$]{0.5}(86,58)(81,64)
\dottedline[\href{http://news.google.co.uk/archivesearch?as_q=+Thomas+Schieffer+Taro+Aso&num=10&hl=en&btnG=Search+Archives&as_epq=&as_oq=&as_eq=&as_user_ldate=2005&as_user_hdate=2007&lr=&as_scoring=t}{$\cdot$}]{3}(116,80)(118,74)
\dottedline[$\bullet$]{0.5}(116,80)(118,74)
\dottedline[\href{http://news.google.co.uk/archivesearch?as_q=Samir+Geagea+Omar+Karami&num=10&hl=en&btnG=Search+Archives&as_epq=&as_oq=&as_eq=&as_user_ldate=2005&as_user_hdate=2007&lr=&as_scoring=t}{$\cdot$}]{3}(70,7)(76,7)
\dottedline[$\bullet$]{0.5}(70,7)(76,7)
\dottedline[\href{http://news.google.co.uk/archivesearch?as_q=L\'{a}szl\'{o}+S\'{o}lyom+Laura+Bush&num=10&hl=en&btnG=Search+Archives&as_epq=&as_oq=&as_eq=&as_user_ldate=2005&as_user_hdate=2007&lr=&as_scoring=t}{$\cdot$}]{3}(47,118)(54,119)
\dottedline[$\bullet$]{0.5}(47,118)(54,119)
\dottedline[\href{http://news.google.co.uk/archivesearch?as_q=Brad+Berenson+Vaughn+Walker&num=10&hl=en&btnG=Search+Archives&as_epq=&as_oq=&as_eq=&as_user_ldate=2005&as_user_hdate=2007&lr=&as_scoring=t}{$\cdot$}]{3}(98,49)(97,43)
\dottedline[$\bullet$]{0.5}(98,49)(97,43)
\dottedline[\href{http://news.google.co.uk/archivesearch?as_q=Rangin+Dadfar+Spanta+Pervez+Musharraf&num=10&hl=en&btnG=Search+Archives&as_epq=&as_oq=&as_eq=&as_user_ldate=2005&as_user_hdate=2007&lr=&as_scoring=t}{$\cdot$}]{3}(65,111)(66,103)
\dottedline[$\bullet$]{0.5}(65,111)(66,103)
\dottedline[\href{http://news.google.co.uk/archivesearch?as_q=Riek+Machar+Joseph+Kony&num=10&hl=en&btnG=Search+Archives&as_epq=&as_oq=&as_eq=&as_user_ldate=2005&as_user_hdate=2007&lr=&as_scoring=t}{$\cdot$}]{3}(37,61)(37,54)
\dottedline[$\bullet$]{0.5}(37,61)(37,54)
\dottedline[\href{http://news.google.co.uk/archivesearch?as_q=Tony+Snow+Norman+Mineta&num=10&hl=en&btnG=Search+Archives&as_epq=&as_oq=&as_eq=&as_user_ldate=2005&as_user_hdate=2007&lr=&as_scoring=t}{$\cdot$}]{3}(98,74)(94,81)
\dottedline[$\bullet$]{0.5}(98,74)(94,81)
\color{red}
\dottedline[\href{http://news.google.co.uk/archivesearch?as_q=George+W.+Bush+Michael+Bloomberg&num=10&hl=en&btnG=Search+Archives&as_epq=&as_oq=&as_eq=&as_user_ldate=2005&as_user_hdate=2007&lr=&as_scoring=t}{$\cdot$}]{3}(91,91)(100,87)
\dottedline[$\bullet$]{0.5}(91,91)(100,87)
\dottedline[\href{http://news.google.co.uk/archivesearch?as_q=Kofi+Annan+Condoleezza+Rice&num=10&hl=en&btnG=Search+Archives&as_epq=&as_oq=&as_eq=&as_user_ldate=2005&as_user_hdate=2007&lr=&as_scoring=t}{$\cdot$}]{3}(65,81)(69,92)
\dottedline[$\bullet$]{0.5}(65,81)(69,92)
\dottedline[\href{http://news.google.co.uk/archivesearch?as_q=Kofi+Annan+Tzipi+Livni&num=10&hl=en&btnG=Search+Archives&as_epq=&as_oq=&as_eq=&as_user_ldate=2005&as_user_hdate=2007&lr=&as_scoring=t}{$\cdot$}]{3}(65,81)(56,89)
\dottedline[$\bullet$]{0.5}(65,81)(56,89)
\dottedline[\href{http://news.google.co.uk/archivesearch?as_q=Donald+Rumsfeld+George+Casey&num=10&hl=en&btnG=Search+Archives&as_epq=&as_oq=&as_eq=&as_user_ldate=2005&as_user_hdate=2007&lr=&as_scoring=t}{$\cdot$}]{3}(18,26)(22,31)
\dottedline[$\bullet$]{0.5}(18,26)(22,31)
\dottedline[\href{http://news.google.co.uk/archivesearch?as_q=Condoleezza+Rice+Kofi+Annan&num=10&hl=en&btnG=Search+Archives&as_epq=&as_oq=&as_eq=&as_user_ldate=2005&as_user_hdate=2007&lr=&as_scoring=t}{$\cdot$}]{3}(69,92)(65,81)
\dottedline[$\bullet$]{0.5}(69,92)(65,81)
\dottedline[\href{http://news.google.co.uk/archivesearch?as_q=Condoleezza+Rice+Pervez+Musharraf&num=10&hl=en&btnG=Search+Archives&as_epq=&as_oq=&as_eq=&as_user_ldate=2005&as_user_hdate=2007&lr=&as_scoring=t}{$\cdot$}]{3}(69,92)(66,103)
\dottedline[$\bullet$]{0.5}(69,92)(66,103)
\dottedline[\href{http://news.google.co.uk/archivesearch?as_q=Condoleezza+Rice+Hamid+Karzai&num=10&hl=en&btnG=Search+Archives&as_epq=&as_oq=&as_eq=&as_user_ldate=2005&as_user_hdate=2007&lr=&as_scoring=t}{$\cdot$}]{3}(69,92)(80,100)
\dottedline[$\bullet$]{0.5}(69,92)(80,100)
\dottedline[\href{http://news.google.co.uk/archivesearch?as_q=Condoleezza+Rice+Tzipi+Livni&num=10&hl=en&btnG=Search+Archives&as_epq=&as_oq=&as_eq=&as_user_ldate=2005&as_user_hdate=2007&lr=&as_scoring=t}{$\cdot$}]{3}(69,92)(56,89)
\dottedline[$\bullet$]{0.5}(69,92)(56,89)
\dottedline[\href{http://news.google.co.uk/archivesearch?as_q=Condoleezza+Rice+Emile+Lahoud&num=10&hl=en&btnG=Search+Archives&as_epq=&as_oq=&as_eq=&as_user_ldate=2005&as_user_hdate=2007&lr=&as_scoring=t}{$\cdot$}]{3}(69,92)(80,89)
\dottedline[$\bullet$]{0.5}(69,92)(80,89)
\dottedline[\href{http://news.google.co.uk/archivesearch?as_q=Pervez+Musharraf+Condoleezza+Rice&num=10&hl=en&btnG=Search+Archives&as_epq=&as_oq=&as_eq=&as_user_ldate=2005&as_user_hdate=2007&lr=&as_scoring=t}{$\cdot$}]{3}(66,103)(69,92)
\dottedline[$\bullet$]{0.5}(66,103)(69,92)
\dottedline[\href{http://news.google.co.uk/archivesearch?as_q=John+Howard+Philip+Ruddock&num=10&hl=en&btnG=Search+Archives&as_epq=&as_oq=&as_eq=&as_user_ldate=2005&as_user_hdate=2007&lr=&as_scoring=t}{$\cdot$}]{3}(36,28)(35,20)
\dottedline[$\bullet$]{0.5}(36,28)(35,20)
\dottedline[\href{http://news.google.co.uk/archivesearch?as_q=Yasser+Arafat+Ehud+Barak&num=10&hl=en&btnG=Search+Archives&as_epq=&as_oq=&as_eq=&as_user_ldate=2005&as_user_hdate=2007&lr=&as_scoring=t}{$\cdot$}]{3}(106,62)(112,64)
\dottedline[$\bullet$]{0.5}(106,62)(112,64)
\dottedline[\href{http://news.google.co.uk/archivesearch?as_q=Hamid+Karzai+Condoleezza+Rice&num=10&hl=en&btnG=Search+Archives&as_epq=&as_oq=&as_eq=&as_user_ldate=2005&as_user_hdate=2007&lr=&as_scoring=t}{$\cdot$}]{3}(80,100)(69,92)
\dottedline[$\bullet$]{0.5}(80,100)(69,92)
\dottedline[\href{http://news.google.co.uk/archivesearch?as_q=Hamid+Karzai+Mullah+Omar&num=10&hl=en&btnG=Search+Archives&as_epq=&as_oq=&as_eq=&as_user_ldate=2005&as_user_hdate=2007&lr=&as_scoring=t}{$\cdot$}]{3}(80,100)(77,109)
\dottedline[$\bullet$]{0.5}(80,100)(77,109)
\dottedline[\href{http://news.google.co.uk/archivesearch?as_q=Tassos+Papadopoulos+Mehmet+Ali+Talat&num=10&hl=en&btnG=Search+Archives&as_epq=&as_oq=&as_eq=&as_user_ldate=2005&as_user_hdate=2007&lr=&as_scoring=t}{$\cdot$}]{3}(56,5)(57,11)
\dottedline[$\bullet$]{0.5}(56,5)(57,11)
\dottedline[\href{http://news.google.co.uk/archivesearch?as_q=Ehud+Olmert+Mahmoud+Abbas&num=10&hl=en&btnG=Search+Archives&as_epq=&as_oq=&as_eq=&as_user_ldate=2005&as_user_hdate=2007&lr=&as_scoring=t}{$\cdot$}]{3}(41,106)(43,96)
\dottedline[$\bullet$]{0.5}(41,106)(43,96)
\dottedline[\href{http://news.google.co.uk/archivesearch?as_q=Mullah+Omar+Hamid+Karzai&num=10&hl=en&btnG=Search+Archives&as_epq=&as_oq=&as_eq=&as_user_ldate=2005&as_user_hdate=2007&lr=&as_scoring=t}{$\cdot$}]{3}(77,109)(80,100)
\dottedline[$\bullet$]{0.5}(77,109)(80,100)
\dottedline[\href{http://news.google.co.uk/archivesearch?as_q=Michael+Bloomberg+George+W.+Bush&num=10&hl=en&btnG=Search+Archives&as_epq=&as_oq=&as_eq=&as_user_ldate=2005&as_user_hdate=2007&lr=&as_scoring=t}{$\cdot$}]{3}(100,87)(91,91)
\dottedline[$\bullet$]{0.5}(100,87)(91,91)
\dottedline[\href{http://news.google.co.uk/archivesearch?as_q=Abu+Musab+al-Zarqawi+Ayman+al-Zawahiri&num=10&hl=en&btnG=Search+Archives&as_epq=&as_oq=&as_eq=&as_user_ldate=2005&as_user_hdate=2007&lr=&as_scoring=t}{$\cdot$}]{3}(104,97)(93,98)
\dottedline[$\bullet$]{0.5}(104,97)(93,98)
\dottedline[\href{http://news.google.co.uk/archivesearch?as_q=Shimon+Peres+Mahmoud+Abbas&num=10&hl=en&btnG=Search+Archives&as_epq=&as_oq=&as_eq=&as_user_ldate=2005&as_user_hdate=2007&lr=&as_scoring=t}{$\cdot$}]{3}(49,102)(43,96)
\dottedline[$\bullet$]{0.5}(49,102)(43,96)
\dottedline[\href{http://news.google.co.uk/archivesearch?as_q=Mehmet+Ali+Talat+Tassos+Papadopoulos&num=10&hl=en&btnG=Search+Archives&as_epq=&as_oq=&as_eq=&as_user_ldate=2005&as_user_hdate=2007&lr=&as_scoring=t}{$\cdot$}]{3}(57,11)(56,5)
\dottedline[$\bullet$]{0.5}(57,11)(56,5)
\dottedline[\href{http://news.google.co.uk/archivesearch?as_q=George+Casey+Donald+Rumsfeld&num=10&hl=en&btnG=Search+Archives&as_epq=&as_oq=&as_eq=&as_user_ldate=2005&as_user_hdate=2007&lr=&as_scoring=t}{$\cdot$}]{3}(22,31)(18,26)
\dottedline[$\bullet$]{0.5}(22,31)(18,26)
\dottedline[\href{http://news.google.co.uk/archivesearch?as_q=Philip+Ruddock+John+Howard&num=10&hl=en&btnG=Search+Archives&as_epq=&as_oq=&as_eq=&as_user_ldate=2005&as_user_hdate=2007&lr=&as_scoring=t}{$\cdot$}]{3}(35,20)(36,28)
\dottedline[$\bullet$]{0.5}(35,20)(36,28)
\dottedline[\href{http://news.google.co.uk/archivesearch?as_q=Jalal+Talabani+Mahmoud+Ahmadinejad&num=10&hl=en&btnG=Search+Archives&as_epq=&as_oq=&as_eq=&as_user_ldate=2005&as_user_hdate=2007&lr=&as_scoring=t}{$\cdot$}]{3}(86,17)(81,21)
\dottedline[$\bullet$]{0.5}(86,17)(81,21)
\dottedline[\href{http://news.google.co.uk/archivesearch?as_q=Rajiv+Gandhi+Indira+Gandhi&num=10&hl=en&btnG=Search+Archives&as_epq=&as_oq=&as_eq=&as_user_ldate=2005&as_user_hdate=2007&lr=&as_scoring=t}{$\cdot$}]{3}(67,38)(65,44)
\dottedline[$\bullet$]{0.5}(67,38)(65,44)
\dottedline[\href{http://news.google.co.uk/archivesearch?as_q=Ehud+Barak+Yasser+Arafat&num=10&hl=en&btnG=Search+Archives&as_epq=&as_oq=&as_eq=&as_user_ldate=2005&as_user_hdate=2007&lr=&as_scoring=t}{$\cdot$}]{3}(112,64)(106,62)
\dottedline[$\bullet$]{0.5}(112,64)(106,62)
\dottedline[\href{http://news.google.co.uk/archivesearch?as_q=Mahmoud+Abbas+Ehud+Olmert&num=10&hl=en&btnG=Search+Archives&as_epq=&as_oq=&as_eq=&as_user_ldate=2005&as_user_hdate=2007&lr=&as_scoring=t}{$\cdot$}]{3}(43,96)(41,106)
\dottedline[$\bullet$]{0.5}(43,96)(41,106)
\dottedline[\href{http://news.google.co.uk/archivesearch?as_q=Mahmoud+Abbas+Shimon+Peres&num=10&hl=en&btnG=Search+Archives&as_epq=&as_oq=&as_eq=&as_user_ldate=2005&as_user_hdate=2007&lr=&as_scoring=t}{$\cdot$}]{3}(43,96)(49,102)
\dottedline[$\bullet$]{0.5}(43,96)(49,102)
\dottedline[\href{http://news.google.co.uk/archivesearch?as_q=Mahmoud+Abbas+Ismail+Haniya&num=10&hl=en&btnG=Search+Archives&as_epq=&as_oq=&as_eq=&as_user_ldate=2005&as_user_hdate=2007&lr=&as_scoring=t}{$\cdot$}]{3}(43,96)(33,94)
\dottedline[$\bullet$]{0.5}(43,96)(33,94)
\dottedline[\href{http://news.google.co.uk/archivesearch?as_q=Tzipi+Livni+Kofi+Annan&num=10&hl=en&btnG=Search+Archives&as_epq=&as_oq=&as_eq=&as_user_ldate=2005&as_user_hdate=2007&lr=&as_scoring=t}{$\cdot$}]{3}(56,89)(65,81)
\dottedline[$\bullet$]{0.5}(56,89)(65,81)
\dottedline[\href{http://news.google.co.uk/archivesearch?as_q=Tzipi+Livni+Condoleezza+Rice&num=10&hl=en&btnG=Search+Archives&as_epq=&as_oq=&as_eq=&as_user_ldate=2005&as_user_hdate=2007&lr=&as_scoring=t}{$\cdot$}]{3}(56,89)(69,92)
\dottedline[$\bullet$]{0.5}(56,89)(69,92)
\dottedline[\href{http://news.google.co.uk/archivesearch?as_q=Ayman+al-Zawahiri+Abu+Musab+al-Zarqawi&num=10&hl=en&btnG=Search+Archives&as_epq=&as_oq=&as_eq=&as_user_ldate=2005&as_user_hdate=2007&lr=&as_scoring=t}{$\cdot$}]{3}(93,98)(104,97)
\dottedline[$\bullet$]{0.5}(93,98)(104,97)
\dottedline[\href{http://news.google.co.uk/archivesearch?as_q=Emile+Lahoud+Condoleezza+Rice&num=10&hl=en&btnG=Search+Archives&as_epq=&as_oq=&as_eq=&as_user_ldate=2005&as_user_hdate=2007&lr=&as_scoring=t}{$\cdot$}]{3}(80,89)(69,92)
\dottedline[$\bullet$]{0.5}(80,89)(69,92)
\dottedline[\href{http://news.google.co.uk/archivesearch?as_q=Indira+Gandhi+Rajiv+Gandhi&num=10&hl=en&btnG=Search+Archives&as_epq=&as_oq=&as_eq=&as_user_ldate=2005&as_user_hdate=2007&lr=&as_scoring=t}{$\cdot$}]{3}(65,44)(67,38)
\dottedline[$\bullet$]{0.5}(65,44)(67,38)
\dottedline[\href{http://news.google.co.uk/archivesearch?as_q=Ismail+Haniya+Mahmoud+Abbas&num=10&hl=en&btnG=Search+Archives&as_epq=&as_oq=&as_eq=&as_user_ldate=2005&as_user_hdate=2007&lr=&as_scoring=t}{$\cdot$}]{3}(33,94)(43,96)
\dottedline[$\bullet$]{0.5}(33,94)(43,96)
\dottedline[\href{http://news.google.co.uk/archivesearch?as_q=Mahmoud+Ahmadinejad+Jalal+Talabani&num=10&hl=en&btnG=Search+Archives&as_epq=&as_oq=&as_eq=&as_user_ldate=2005&as_user_hdate=2007&lr=&as_scoring=t}{$\cdot$}]{3}(81,21)(86,17)
\dottedline[$\bullet$]{0.5}(81,21)(86,17)
\color{blue}
\put(91,91){\circle*{2}}
\color{black}
\put(91,92){\href{http://en.wikipedia.org/wiki/George W. Bush}{\textit{\textbf{ George W. Bush}}}}
\color{blue}
\put(13,80){\circle*{2}}
\color{black}
\put(13,81){\href{http://en.wikipedia.org/wiki/Tony Blair}{\textit{\textbf{ Tony Blair}}}}
\color{red}
\put(65,81){\circle*{2}}
\color{red}
\put(65,82){\href{http://en.wikipedia.org/wiki/Kofi Annan}{\textit{\textbf{ Kofi Annan}}}}
\color{blue}
\put(20,65){\circle*{2}}
\color{black}
\put(20,66){\href{http://en.wikipedia.org/wiki/Dick Cheney}{\textit{\textbf{ Dick Cheney}}}}
\color{blue}
\put(18,26){\circle*{2}}
\color{black}
\put(18,27){\href{http://en.wikipedia.org/wiki/Donald Rumsfeld}{\textit{\textbf{ Donald Rumsfeld}}}}
\color{red}
\put(69,92){\circle*{2}}
\color{red}
\put(69,93){\href{http://en.wikipedia.org/wiki/Condoleezza Rice}{\textit{\textbf{ Condoleezza Rice}}}}
\color{blue}
\put(66,103){\circle*{2}}
\color{black}
\put(66,104){\href{http://en.wikipedia.org/wiki/Pervez Musharraf}{\textit{\textbf{ Pervez Musharraf}}}}
\color{blue}
\put(36,28){\circle*{2}}
\color{black}
\put(36,29){\href{http://en.wikipedia.org/wiki/John Howard}{\textit{\textbf{ John Howard}}}}
\color{blue}
\put(89,103){\circle*{2}}
\color{black}
\put(89,104){\href{http://en.wikipedia.org/wiki/Junichiro Koizumi}{\textit{\textbf{ Junichiro Koizumi}}}}
\color{blue}
\put(115,48){\circle*{2}}
\color{black}
\put(115,49){\href{http://en.wikipedia.org/wiki/Gordon Brown}{\textit{\textbf{ Gordon Brown}}}}
\color{blue}
\put(106,62){\circle*{2}}
\color{black}
\put(106,63){\href{http://en.wikipedia.org/wiki/Yasser Arafat}{\textit{\textbf{ Yasser Arafat}}}}
\color{red}
\put(80,100){\circle*{2}}
\color{red}
\put(80,101){\href{http://en.wikipedia.org/wiki/Hamid Karzai}{\textit{\textbf{ Hamid Karzai}}}}
\color{blue}
\put(38,87){\circle*{2}}
\color{black}
\put(38,88){\href{http://en.wikipedia.org/wiki/Hosni Mubarak}{\textit{\textbf{ Hosni Mubarak}}}}
\color{blue}
\put(49,82){\circle*{2}}
\color{black}
\put(49,83){\href{http://en.wikipedia.org/wiki/Javier Solana}{\textit{\textbf{ Javier Solana}}}}
\color{blue}
\put(12,87){\circle*{2}}
\color{black}
\put(12,88){\href{http://en.wikipedia.org/wiki/John Reid}{\textit{\textbf{ John Reid}}}}
\color{blue}
\put(81,64){\circle*{2}}
\color{black}
\put(81,65){\href{http://en.wikipedia.org/wiki/Abdullah G\"{u}l}{\textit{\textbf{ Abdullah G\"{u}l}}}}
\color{blue}
\put(29,17){\circle*{2}}
\color{black}
\put(29,18){\href{http://en.wikipedia.org/wiki/Alexander Downer}{\textit{\textbf{ Alexander Downer}}}}
\color{red}
\put(56,5){\circle*{2}}
\color{red}
\put(56,6){\href{http://en.wikipedia.org/wiki/Tassos Papadopoulos}{\textit{\textbf{ Tassos Papadopoulos}}}}
\color{blue}
\put(8,75){\circle*{2}}
\color{black}
\put(8,76){\href{http://en.wikipedia.org/wiki/Vojislav Koatunica}{\textit{\textbf{ Vojislav Koatunica}}}}
\color{blue}
\put(18,96){\circle*{2}}
\color{black}
\put(18,97){\href{http://en.wikipedia.org/wiki/Ali Khamenei}{\textit{\textbf{ Ali Khamenei}}}}
\color{blue}
\put(86,34){\circle*{2}}
\color{black}
\put(86,35){\href{http://en.wikipedia.org/wiki/Manmohan Singh}{\textit{\textbf{ Manmohan Singh}}}}
\color{red}
\put(60,24){\circle*{2}}
\color{red}
\put(60,25){\href{http://en.wikipedia.org/wiki/Robert Mueller}{\textit{\textbf{ Robert Mueller}}}}
\color{red}
\put(41,106){\circle*{2}}
\color{red}
\put(41,107){\href{http://en.wikipedia.org/wiki/Ehud Olmert}{\textit{\textbf{ Ehud Olmert}}}}
\color{red}
\put(77,109){\circle*{2}}
\color{red}
\put(77,110){\href{http://en.wikipedia.org/wiki/Mullah Omar}{\textit{\textbf{ Mullah Omar}}}}
\color{blue}
\put(100,87){\circle*{2}}
\color{black}
\put(100,88){\href{http://en.wikipedia.org/wiki/Michael Bloomberg}{\textit{\textbf{ Michael Bloomberg}}}}
\color{blue}
\put(104,97){\circle*{2}}
\color{black}
\put(104,98){\href{http://en.wikipedia.org/wiki/Abu Musab al-Zarqawi}{\textit{\textbf{ Abu Musab al-Zarqawi}}}}
\color{red}
\put(49,102){\circle*{2}}
\color{red}
\put(49,103){\href{http://en.wikipedia.org/wiki/Shimon Peres}{\textit{\textbf{ Shimon Peres}}}}
\color{blue}
\put(37,47){\circle*{2}}
\color{black}
\put(37,48){\href{http://en.wikipedia.org/wiki/Yoweri Museveni}{\textit{\textbf{ Yoweri Museveni}}}}
\color{red}
\put(57,11){\circle*{2}}
\color{red}
\put(57,12){\href{http://en.wikipedia.org/wiki/Mehmet Ali Talat}{\textit{\textbf{ Mehmet Ali Talat}}}}
\color{red}
\put(22,31){\circle*{2}}
\color{red}
\put(22,32){\href{http://en.wikipedia.org/wiki/George Casey}{\textit{\textbf{ George Casey}}}}
\color{blue}
\put(54,119){\circle*{2}}
\color{black}
\put(54,120){\href{http://en.wikipedia.org/wiki/Laura Bush}{\textit{\textbf{ Laura Bush}}}}
\color{red}
\put(49,34){\circle*{2}}
\color{red}
\put(49,35){\href{http://en.wikipedia.org/wiki/Joseph Biden}{\textit{\textbf{ Joseph Biden}}}}
\color{red}
\put(35,20){\circle*{2}}
\color{red}
\put(35,21){\href{http://en.wikipedia.org/wiki/Philip Ruddock}{\textit{\textbf{ Philip Ruddock}}}}
\color{blue}
\put(86,17){\circle*{2}}
\color{black}
\put(86,18){\href{http://en.wikipedia.org/wiki/Jalal Talabani}{\textit{\textbf{ Jalal Talabani}}}}
\color{blue}
\put(110,35){\circle*{2}}
\color{black}
\put(110,36){\href{http://en.wikipedia.org/wiki/Des Browne}{\textit{\textbf{ Des Browne}}}}
\color{blue}
\put(88,112){\circle*{2}}
\color{black}
\put(88,113){\href{http://en.wikipedia.org/wiki/Stephen Harper}{\textit{\textbf{ Stephen Harper}}}}
\color{red}
\put(67,38){\circle*{2}}
\color{red}
\put(67,39){\href{http://en.wikipedia.org/wiki/Rajiv Gandhi}{\textit{\textbf{ Rajiv Gandhi}}}}
\color{red}
\put(60,72){\circle*{2}}
\color{red}
\put(60,73){\href{http://en.wikipedia.org/wiki/Omar el-Bashir}{\textit{\textbf{ Omar el-Bashir}}}}
\color{red}
\put(112,64){\circle*{2}}
\color{red}
\put(112,65){\href{http://en.wikipedia.org/wiki/Ehud Barak}{\textit{\textbf{ Ehud Barak}}}}
\color{blue}
\put(43,9){\circle*{2}}
\color{black}
\put(43,10){\href{http://en.wikipedia.org/wiki/ James Jones}{\textit{\textbf{  James Jones}}}}
\color{red}
\put(43,96){\circle*{2}}
\color{red}
\put(43,97){\href{http://en.wikipedia.org/wiki/Mahmoud Abbas}{\textit{\textbf{ Mahmoud Abbas}}}}
\color{blue}
\put(91,13){\circle*{2}}
\color{black}
\put(91,14){\href{http://en.wikipedia.org/wiki/Mahmoud Othman}{\textit{\textbf{ Mahmoud Othman}}}}
\color{blue}
\put(50,41){\circle*{2}}
\color{black}
\put(50,42){\href{http://en.wikipedia.org/wiki/Mark Warner}{\textit{\textbf{ Mark Warner}}}}
\color{red}
\put(113,41){\circle*{2}}
\color{red}
\put(113,42){\href{http://en.wikipedia.org/wiki/Clare Short}{\textit{\textbf{ Clare Short}}}}
\color{red}
\put(56,89){\circle*{2}}
\color{red}
\put(56,90){\href{http://en.wikipedia.org/wiki/Tzipi Livni}{\textit{\textbf{ Tzipi Livni}}}}
\color{blue}
\put(20,103){\circle*{2}}
\color{black}
\put(20,104){\href{http://en.wikipedia.org/wiki/Abdoulaye Wade}{\textit{\textbf{ Abdoulaye Wade}}}}
\color{red}
\put(55,64){\circle*{2}}
\color{red}
\put(55,65){\href{http://en.wikipedia.org/wiki/Alpha Oumar Konare}{\textit{\textbf{ Alpha Oumar Konare}}}}
\color{blue}
\put(119,54){\circle*{2}}
\color{black}
\put(119,55){\href{http://en.wikipedia.org/wiki/Alex Salmond}{\textit{\textbf{ Alex Salmond}}}}
\color{blue}
\put(20,50){\circle*{2}}
\color{black}
\put(20,51){\href{http://en.wikipedia.org/wiki/Khaled Mashal}{\textit{\textbf{ Khaled Mashal}}}}
\color{blue}
\put(97,43){\circle*{2}}
\color{black}
\put(97,44){\href{http://en.wikipedia.org/wiki/Vaughn Walker}{\textit{\textbf{ Vaughn Walker}}}}
\color{blue}
\put(71,61){\circle*{2}}
\color{black}
\put(71,62){\href{http://en.wikipedia.org/wiki/Frank-Walter Steinmeier}{\textit{\textbf{ Frank-Walter Steinmeier}}}}
\color{red}
\put(93,98){\circle*{2}}
\color{red}
\put(93,99){\href{http://en.wikipedia.org/wiki/Ayman al-Zawahiri}{\textit{\textbf{ Ayman al-Zawahiri}}}}
\color{red}
\put(80,89){\circle*{2}}
\color{red}
\put(80,90){\href{http://en.wikipedia.org/wiki/Emile Lahoud}{\textit{\textbf{ Emile Lahoud}}}}
\color{blue}
\put(99,26){\circle*{2}}
\color{black}
\put(99,27){\href{http://en.wikipedia.org/wiki/Amir Peretz}{\textit{\textbf{ Amir Peretz}}}}
\color{blue}
\put(23,44){\circle*{2}}
\color{black}
\put(23,45){\href{http://en.wikipedia.org/wiki/Haim Ramon}{\textit{\textbf{ Haim Ramon}}}}
\color{blue}
\put(7,57){\circle*{2}}
\color{black}
\put(7,58){\href{http://en.wikipedia.org/wiki/Xanana Gusm\H{a}o}{\textit{\textbf{ Xanana Gusm\H{a}o}}}}
\color{red}
\put(5,64){\circle*{2}}
\color{red}
\put(5,65){\href{http://en.wikipedia.org/wiki/Mar\'{\i} Alkatiri}{\textit{\textbf{ Mar\'{\i} Alkatiri}}}}
\color{red}
\put(94,81){\circle*{2}}
\color{red}
\put(94,82){\href{http://en.wikipedia.org/wiki/Norman Mineta}{\textit{\textbf{ Norman Mineta}}}}
\color{red}
\put(42,16){\circle*{2}}
\color{red}
\put(42,17){\href{http://en.wikipedia.org/wiki/Brendan Nelson}{\textit{\textbf{ Brendan Nelson}}}}
\color{blue}
\put(7,45){\circle*{2}}
\color{black}
\put(7,46){\href{http://en.wikipedia.org/wiki/David Cameron}{\textit{\textbf{ David Cameron}}}}
\color{red}
\put(37,54){\circle*{2}}
\color{red}
\put(37,55){\href{http://en.wikipedia.org/wiki/Joseph Kony}{\textit{\textbf{ Joseph Kony}}}}
\color{blue}
\put(10,38){\circle*{2}}
\color{black}
\put(10,39){\href{http://en.wikipedia.org/wiki/Shami Chakrabarti}{\textit{\textbf{ Shami Chakrabarti}}}}
\color{blue}
\put(81,39){\circle*{2}}
\color{black}
\put(81,40){\href{http://en.wikipedia.org/wiki/Mangala Samaraweera}{\textit{\textbf{ Mangala Samaraweera}}}}
\color{blue}
\put(118,74){\circle*{2}}
\color{black}
\put(118,75){\href{http://en.wikipedia.org/wiki/Taro Aso}{\textit{\textbf{ Taro Aso}}}}
\color{blue}
\put(65,44){\circle*{2}}
\color{black}
\put(65,45){\href{http://en.wikipedia.org/wiki/Indira Gandhi}{\textit{\textbf{ Indira Gandhi}}}}
\color{blue}
\put(66,23){\circle*{2}}
\color{black}
\put(66,24){\href{http://en.wikipedia.org/wiki/Larry King}{\textit{\textbf{ Larry King}}}}
\color{blue}
\put(35,102){\circle*{2}}
\color{black}
\put(35,103){\href{http://en.wikipedia.org/wiki/Elie Wiesel}{\textit{\textbf{ Elie Wiesel}}}}
\color{red}
\put(104,23){\circle*{2}}
\color{red}
\put(104,24){\href{http://en.wikipedia.org/wiki/Dan Halutz}{\textit{\textbf{ Dan Halutz}}}}
\color{red}
\put(33,94){\circle*{2}}
\color{red}
\put(33,95){\href{http://en.wikipedia.org/wiki/Ismail Haniya}{\textit{\textbf{ Ismail Haniya}}}}
\color{red}
\put(24,71){\circle*{2}}
\color{red}
\put(24,72){\href{http://en.wikipedia.org/wiki/John King}{\textit{\textbf{ John King}}}}
\color{red}
\put(76,7){\circle*{2}}
\color{red}
\put(76,8){\href{http://en.wikipedia.org/wiki/Omar Karami}{\textit{\textbf{ Omar Karami}}}}
\color{red}
\put(72,70){\circle*{2}}
\color{red}
\put(72,71){\href{http://en.wikipedia.org/wiki/Manouchehr Mottaki}{\textit{\textbf{ Manouchehr Mottaki}}}}
\color{blue}
\put(100,105){\circle*{2}}
\color{black}
\put(100,106){\href{http://en.wikipedia.org/wiki/Nouri al-Maliki}{\textit{\textbf{ Nouri al-Maliki}}}}
\color{blue}
\put(41,80){\circle*{2}}
\color{black}
\put(41,81){\href{http://en.wikipedia.org/wiki/Ali Larijani}{\textit{\textbf{ Ali Larijani}}}}
\color{blue}
\put(81,21){\circle*{2}}
\color{black}
\put(81,22){\href{http://en.wikipedia.org/wiki/Mahmoud Ahmadinejad}{\textit{\textbf{ Mahmoud Ahmadinejad}}}}
\color{red}
\put(86,58){\circle*{2}}
\color{red}
\put(86,59){\href{http://en.wikipedia.org/wiki/Sergey Kislyak}{\textit{\textbf{ Sergey Kislyak}}}}
\color{blue}
\put(116,80){\circle*{2}}
\color{black}
\put(116,81){\href{http://en.wikipedia.org/wiki/ Thomas Schieffer}{\textit{\textbf{  Thomas Schieffer}}}}
\color{blue}
\put(70,7){\circle*{2}}
\color{black}
\put(70,8){\href{http://en.wikipedia.org/wiki/Samir Geagea}{\textit{\textbf{ Samir Geagea}}}}
\color{blue}
\put(47,118){\circle*{2}}
\color{black}
\put(47,119){\href{http://en.wikipedia.org/wiki/L\'{a}szl\'{o} S\'{o}lyom}{\textit{\textbf{ L\'{a}szl\'{o} S\'{o}lyom}}}}
\color{blue}
\put(98,49){\circle*{2}}
\color{black}
\put(98,50){\href{http://en.wikipedia.org/wiki/Brad Berenson}{\textit{\textbf{ Brad Berenson}}}}
\color{blue}
\put(65,111){\circle*{2}}
\color{black}
\put(65,112){\href{http://en.wikipedia.org/wiki/Rangin Dadfar Spanta}{\textit{\textbf{ Rangin Dadfar Spanta}}}}
\color{blue}
\put(37,61){\circle*{2}}
\color{black}
\put(37,62){\href{http://en.wikipedia.org/wiki/Riek Machar}{\textit{\textbf{ Riek Machar}}}}
\color{blue}
\put(98,74){\circle*{2}}
\color{black}
\put(98,75){\href{http://en.wikipedia.org/wiki/Tony Snow}{\textit{\textbf{ Tony Snow}}}}
\end{picture}}\vspace*{-13pt}

\caption{Network of all active individuals for the week ending 28 June
2006. Anomalous links (under sequential Dirichlet process analysis)
and individuals are highlighted in red. The nodes and links of this
graph are interactive in the online version of this paper.}\vspace*{-4pt}
\label{fig:full_network_280606}
\end{figure}
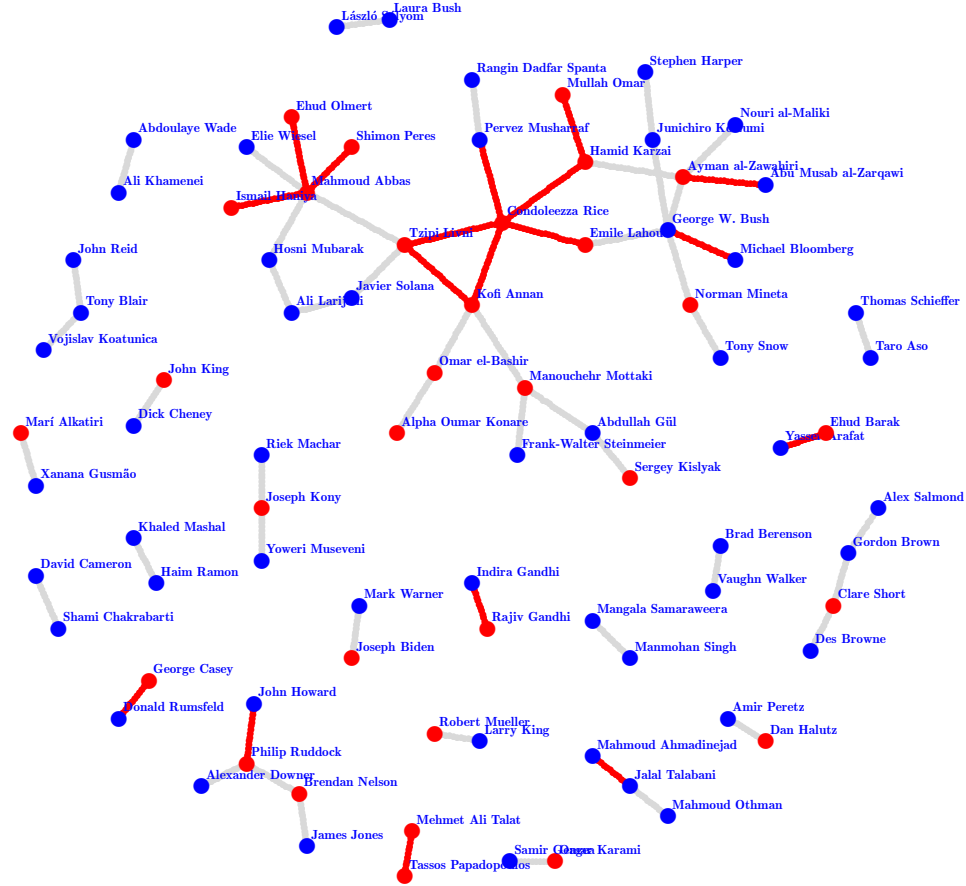

Across parametric and nonparametric models, the highest count of
anomalous nodes identified through either individual behavior or
pairwise interactions occurs on the 73rd week
after the downtime, ending 28 June 2006 (see Figure
\ref{fig:anomaly_counts_hpdp}). Interestingly, this was a time of
continuing violence in the Middle East, which a fortnight later would
see the beginning of the the 2006 Lebanon War; it was also a week in
which the Sudanese government was in disagreement with the United
Nations over humanitarian involvement in the conflict in Darfur. The
extended network from spectral clustering of all communicators in that
week is shown in Figure \ref{fig:full_network_280606}, and important
figures from the stories mentioned can be seen toward the top of the
network. Nodes or links identified as anomalous, according to the
individual or pairwise Poisson--gamma analyses respectively, are
highlighted in red.

\begin{figure}

\includegraphics{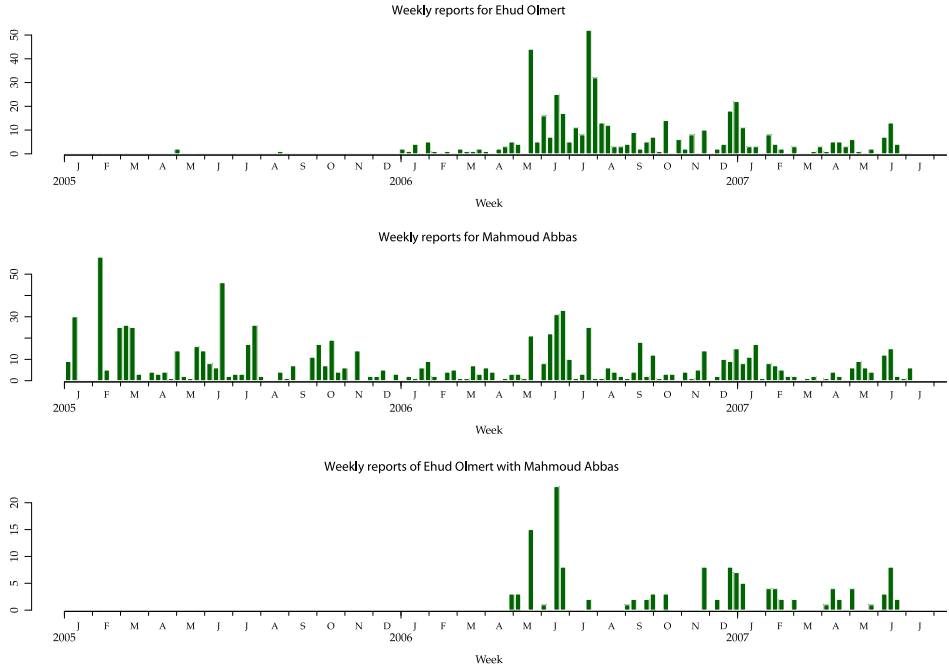}

\caption{Individual news report frequencies for Ehud Olmert and
Mahmoud Abbas, followed by reported contacts between the pair.}
\label{fig:olmert_abbas}
\end{figure}

Taking an example pair of anomalous behaving nodes from the network in
Figure \ref{fig:full_network_280606}, in Figure \ref{fig:olmert_abbas}
we briefly examine the relationship of Ehud Olmert and Mahmoud
Abbas. Mr. Olmert became Prime Minister of Israel on 4 January 2006
(this is apparent from his growing profile in the top panel of Figure
\ref{fig:olmert_abbas}), and Mr. Abbas has been President of the
Palestinian National Authority from 15 January 2005. In the
interesting week ending 28 June 2006, both individuals are showing
higher than usual individual activity (although not their highest
ever, see the top two panels of Figure \ref{fig:olmert_abbas}); but
more significantly, when viewed as a pair (bottom panel of Figure
\ref{fig:olmert_abbas}) they are showing a high peak of connectivity
in the week ending 28 June 2006 which is unmatched at any other time
over the observation period; during that week, they had met in person
for the first time since Mr. Olmert had taken office, and had agreed to
a first official summit. Clearly this behavior should be considered
anomalous and, thus, they take their place in the larger network of
interesting nodes in Figure \ref{fig:full_network_280606}.

\subsection{VAST 2008}

\begin{figure}

\includegraphics{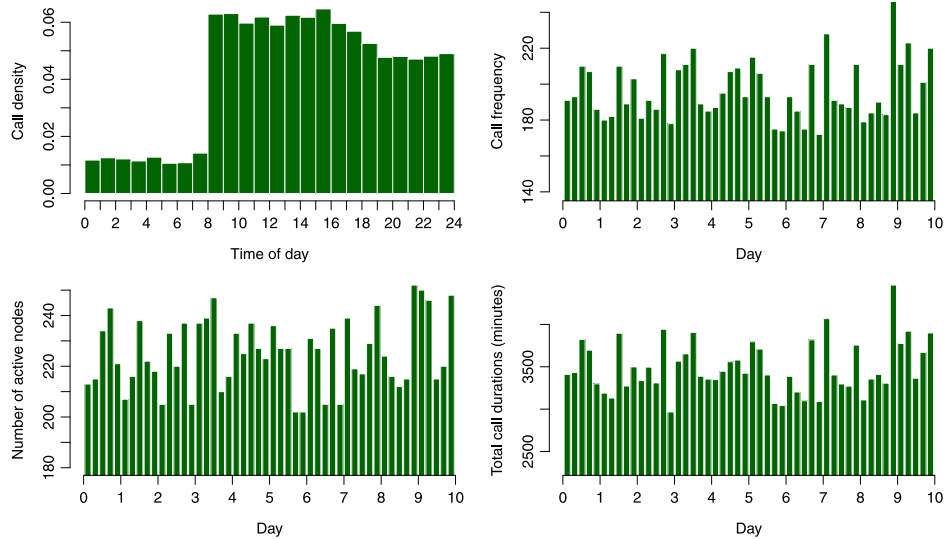}

\caption{Top left: Distribution of phone call start times throughout the
day across the whole network for the VAST 2008 data set. Top right: Call
counts across the whole network after days have been split into five
homogeneous intervals.
Bottom left: Number of active nodes in each subinterval.
Bottom right: Total call durations across the network in each subinterval.}
\label{fig:vast2008_call_time_histograms}
\end{figure}

The simulated call data from the IEEE VAST 2008 Challenge are recorded
in real time, and have sufficient realism that phone calls are not
seen to be made uniformly throughout the day, rather there are peak
and off-peak periods. With only ten days of data, to avoid modeling
of daily cyclical effects and obtain a sufficient number of
homogeneous calling periods for analysis, we used the histogram of daily
phone calls over the ten days across the whole of the network (Figure
\ref{fig:vast2008_call_time_histograms}, top left) to identify the
broad phone
call pattern; the day was then broken up into five subintervals of
equal call frequencies with respect to the histogram. Thus, we obtain
fifty relatively homogeneous periods for our analysis (Figure
\ref{fig:vast2008_call_time_histograms}, top right).

From the results of \citet{yevast2008} we should suspect that the
major anomalous activity begins on the eighth day. This is not
apparent from Figure \ref{fig:vast2008_call_time_histograms}, which on
the bottom row also shows the number of active nodes making or
receiving calls in each period and the total minutes called across the
network in each period. Together, these plots do not reveal any
departures from normality until the end of the ninth day. Clearly
then, from a perspective of timely anomaly detection, this anomalous
behavior is not going to simply coincide with just a change in the
overall activity of the network. It can be noted here that an
improvement offered by our following sequential analysis over the
methods used by \citet{yevast2008} is that the latter made use of the
data across all of the ten days to detect the major change in the
social network; our aim will be to detect the anomaly in (discretized)
real time.

For these data it seems appropriate to use the Markov formulation
given in Equations \eqref{eq:markov_p11} and \eqref{eq:markov_p01};
across all individuals in the data set, the mean empirical estimates
of $\phi_{i.}$ and $\psi_{i.}$ would be 0.63 and 0.48 respectively, so
an individual making calls becomes more likely to continue making
calls into the next period. Because this data set is quite short, we
choose to construct empirical priors using overall means and variances
across all of the call data to get broad insights into typical call
volumes and their variability, which would hopefully mirror the type
of prior knowledge which should be available from a domain
expert. Similar but not identical results can still be achieved with
uninformative priors.

\begin{figure}[b]

\includegraphics{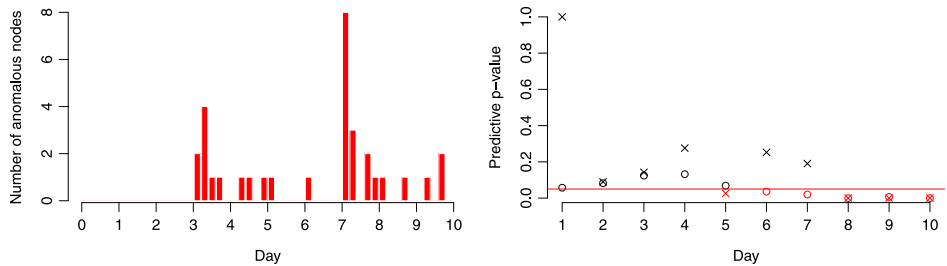}

\caption{Left: The number of anomalous caller IDs found in each time segment
under a Markov--Bernoulli model. Right: The $p$-values of the anomalous
ID cell-tower usage under the multinomial model; the circular points
are the $p$-values when the number of calls is treated as random, the
crosses consider the number of calls as a known quantity.}
\label{fig:vast2008_anomalous_node_counts}
\end{figure}

Figure \ref{fig:vast2008_anomalous_node_counts}(left) shows the
number of
anomalies we find in each time segment under a sequential individual
analysis of the call network nodes using a simple Bernoulli ``on--off''
view of node activity but incorporating the Markov assumption. There
is a clear maximum at the start of the eighth day. All eight of the
anomalous nodes found are from the list deduced by \citet{yevast2008}
using all of the data and several combined methodologies.

\begin{figure}

\includegraphics{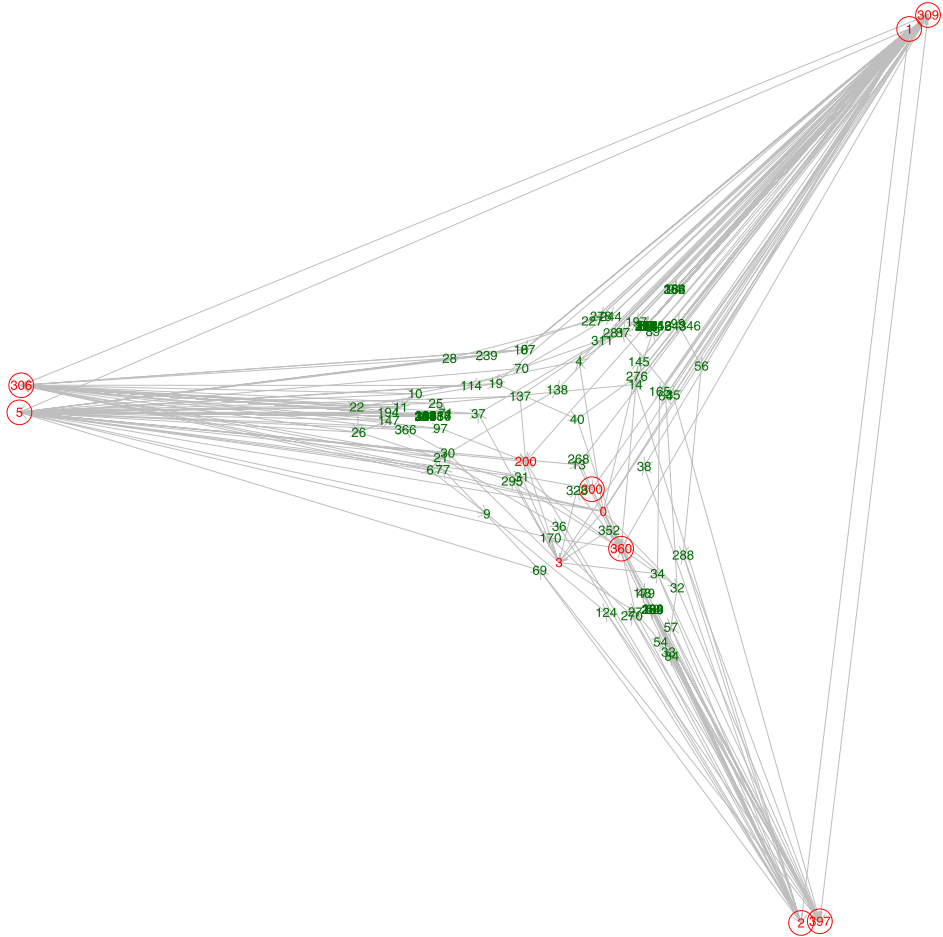}

\caption{Calls between cell phones during the most anomalous period,
occurring at the start of the eighth day. Nodes which were identified
by \protect\citet{yevast2008} as suspicious are colored red; nodes
identified as anomalous in the present analysis are circled.}
\label{fig:vast2008_network}
\end{figure}

A spectral cluster plot made using two components of the symmetric
Laplacian of the historical adjacency matrix [\citet{vonluxburg07}] is
given in Figure \ref{fig:vast2008_network}. The structure is
interesting, with six of the anomalous nodes appearing together in
pairs at extremes of the diagram. As shown in \citet{yevast2008},
these ID pairs are each actually one individual who switches from
using one cell phone to another shortly before the anomalous event
occurs. Besides these three pairs, we find two of the remaining five
individuals declared significant in the social network by
\mbox{\citet{yevast2008}}, although all are very much toward the center of
our filtered graph and so are clearly important figures. Having found
the major anomalous activity, it is a small matter to identify the
remaining participants. Caller ID 200, for example, is the leader of
the social network but is not detected as anomalous by our method;
however, a simple investigation of the call activity of the set of
anomalous nodes detected reveals ID 200 to be the most frequent
communicator in the network with this group, and then the undetected
ID 3 is one of only six nodes who ever communicate with the network
leader; finally, as noted in \citet{yevast2008}, ``the person whose ID
is 0 communicated with all the important people who communicated with
200.''

To better understand the nature of the anomalous behavior of the
circled nodes in Figure \ref{fig:vast2008_network}, we monitor their
collective call activity and cell-tower usage as a group using the
multinomial model from Section \ref{sec:multinomial} with a
Poisson--gamma model for the number of calls; the cell towers reveal
the group's locations on the fictional island each day and so enable
us to track their movements. The group's $p$-values from the multinomial
model for each of the ten days are shown in Figure
\ref{fig:vast2008_anomalous_node_counts}(right), with the two sets of
values corresponding to considering the number of calls made as either
fixed or random. In obtaining these results, a Poisson--gamma(16$/$9, 2$/$9)
model was used for the number of calls so that the expected number of
calls equaled the number of nodes, and a flat Dirichlet $\alpha_k=1/30$
prior used for the multinomial model; see \ref{suppA} [\citet{Heard10}] for details. In
terms of call volumes, the six and seventh days see a big drop in the
group's call activity from an average of over eight calls per day to
just one and two calls in total on those two respective days, followed
by a surge of call activity on the important anomalous eighth day (30
calls) and onward (23 and 27 calls respectively). From a cell tower
perspective, day 2 is found to be fairly anomalous as the group moves
to using a new cell tower, number 30 (three times), whereas day 3 sees
a shift in the balance of how the same set of towers are used. Besides
the high call volumes, the eighth day also sees the group use seven
hitherto unused cell towers---towers~7, 9, 17, 20, 21, 22 and 28, and
the ninth day sees a first use of towers 2 and 12. The predictive
$p$-values drop to near zero on these days.

\section{Discussion}
We have presented a simple statistical framework for monitoring
dynamic social networks, by viewing the frequency of connections
between node pairs as simple counting processes. Bayesian learning of
the distribution of these counts enables predictive $p$-values to be
determined for a new observation. Once a collection of interesting
nodes have been identified in this way, standard network analytical
methods can be used to identify the anomalous network structure.

This methodology has been successfully applied to real and simulated
data sets of moderate size in this paper. Further, it has been
remarked that because the methodology is mostly parallelizable and the
networks are typically sparse, scaling to very large networks is
feasible. For the data sets presented here the analysis is already
fast; for example, for the VAST data set preprocessed into time series
of length 50, identifying anomalous individual activity from the 400
IDs either as callers or through being involved in calls took 2.0
seconds. This timing is based on code run on Matlab 7.3.0 using one
core on a 64-bit 1.86 GHz Xenon quad-core machine.

In both analyses, there was agreement across different count models
about the peak of anomalous behavior. Spectral clustering was used to
identify structure in the anomalous subnetwork, which was in agreement
with knowledge from other sources.

The European Media Monitor data were from a two and a half year
collection period; this could be considered as only a moderate amount
of time in politics, probably overseeing at most one change in
government in any represented country, for example. The simulated cell
phone data covered a very short period, just ten days. The methodology
presented is well suited to short or medium term modeling, as a
global model is fitted across the whole time line and anomalies
detected with respect to this global model.

For a longer term view, a global model is not appropriate, as even
normal behavior between individuals would be expected to evolve. An
adaptive changepoint model with local models fitted within shorter
blocks of time provides a natural extension. Such a model would lie
comfortably within the Bayesian modeling paradigm, although some of
the simplicity of computation enjoyed here would be lost.

\section*{Acknowledgments}
This work was funded by the UK Ministry of Defence Data \& Information
Fusion Defence Technology Centre, project SA012 \textit{Data mining
tools for detecting anomalous clusters in network
communications}. We are most grateful to the EC
Joint Research Centre for kindly providing us with data from their
European Media Monitor.

\begin{supplement}
\sname{Supplement A}\label{suppA}
\stitle{Hurdle exponential family distributions\\}
\slink[doi]{10.1214/10-AOAS329SUPPA}
\slink[url]{http://lib.stat.cmu.edu/aoas/329/supplementA.pdf}
\sdatatype{.pdf}
\sdescription{Details of the Bayesian inferential models considered in
this paper.}
\end{supplement}

\begin{supplement}
\sname{Supplement B}\label{suppB}
\stitle{Matlab/Octave code}
\slink[doi]{10.1214/10-AOAS329SUPPB}
\slink[url]{http://lib.stat.cmu.edu/aoas/329/supplementB.zip}
\sdatatype{.zip}
\sdescription{Matlab code written by DJW for implementing the models
used in this paper.}
\end{supplement}

\printaddresses

\end{document}